\magnification=\magstep1
\hsize=125mm
\overfullrule=0pt
\font\title=cmbx10 scaled\magstep2
\def\R{{\hbox{R}}}

                                               JYFL Preprint 9$\slash$1997

\bigskip

\bigskip

\bigskip

\centerline{\title A Quantum Mechanical Model}

\medskip

\centerline{\title of the}

\medskip

\centerline{\title Reissner-Nordstr\"{o}m Black Hole}

\bigskip

\centerline{Jarmo M\"{a}kel\"{a}\footnote{$^1$}{e-mail: makela@jyfl.jyu.fi}
and Pasi Repo\footnote{$^2$}{e-mail: repo@jyfl.jyu.fi}}

\medskip

\centerline{\it Department of Physics, University of Jyv\"{a}skyl\"{a},
P. O. Box 35, FIN-40351}

\centerline{\it Jyv\"{a}skyl\"{a}, Finland}

\bigskip

\centerline{\bf Abstract}

\medskip

We consider a Hamiltonian quantum theory of spherically symmetric,
asymptotically flat electrovacuum spacetimes. The physical phase space of such spacetimes is
spanned by the mass and the charge parameters $M$ and $Q$ of the
Reissner-Nordstr\"{o}m black hole, together with the corresponding canonical
momenta. In this four-dimensional phase space, we perform a canonical
transformation such that the resulting configuration variables describe the
dynamical properties of Reissner-Nordstr\"{o}m black holes in a natural manner.
The classical Hamiltonian written in terms of these variables and their
conjugate momenta is
replaced by the corresponding self-adjoint Hamiltonian operator, and an
eigenvalue equation for the ADM mass of the hole, from the point of view of a
distant observer at rest, is obtained. Our eigenvalue equation implies that the
ADM mass and the electric charge spectra of the hole are discrete, and the mass
spectrum is bounded below. Moreover, the spectrum of the quantity $M^2-Q^2$ is
strictly positive when an appropriate self-adjoint extension is chosen. The WKB
analysis yields the result that the large eigenvalues of the quantity
$\sqrt{M^2-Q^2}$ are of the form $\sqrt{2n}$, where $n$ is an integer. It turns
out that this result is closely related to Bekenstein's proposal on the
discrete horizon area spectrum of black holes.

\medskip

PACS number(s): 04.60.Ds, 04.20.Fy, 04.60.Kz, 04.70.Dy

\vfill\eject

\centerline{\title 1. Introduction}

\bigskip

\bigskip

    Electrically charged, externally static black holes are very interesting
objects. One of the main interests of these so called Reissner-Nordstr\"{o}m
black holes lies in the peculiar properties of their radiation. For example,
the Hawking temperature of a black hole with mass $M$ and electric charge $Q$
is, in natural units where $\hbar=c=G=k_B=1$[1],
$$
T_H={{\sqrt{M^2-Q^2}}\over{2\pi(M+\sqrt{M^2-Q^2})^2}}.\eqno(1.1)
$$
Hence, we see that when $M^2=Q^2$, i.e. when the hole is extreme, the
temperature of the hole goes to zero, and no radiation comes out from the hole.
Moreover, although an evaluation of the Bekenstein-Hawking entropy of a
non-extreme black hole by means of Euclidean methods implies that
the entropy is exactly one quarter of the area of the apparent horizon of the
hole, the same analysis implies zero entropy for an extreme black hole[2-4].

     What, then, is the difference between extreme and non-extreme black holes?
Of course, one of the fundamental differences lies in their different topological
properties in Euclidean spacetime: the
topology of a non-extreme black hole in Euclidean spacetime is $\Re^2\times
S^2$, whereas that of an extreme black hole is $S^2\times\Re\times S^1$.
However, there is yet another difference which is the primary object of
interest in this
paper. Non-extreme black holes have {\it dynamics}, whereas extreme black holes,
in a certain sense, have not. More precisely, inside the apparent horizon of a
non-extreme black hole there is a region which does not admit a timelike
Killing vector field, from which it follows that it is impossible to choose a
timelike coordinate in such a way that spacetime metric with respect to this
coordinate would be static everywhere in that region. However, an extreme black
hole is static everywhere in its interior as well as in its exterior regions,
which means that everywhere in these domains there is a timelike Killing vector
field, which is orthogonal to a spacelike hypersurface of spacetime. Naively,
this can be seen by considering the spacetime metric of an extreme
Reissner-Nordstr\"{o}m hole, written in the curvature coordinates $T$ and $R$
as:
$$
ds^2 = -\biggl(1-{M\over R}\biggr)^2\,dT^2 + {{dR^2}\over{\biggl(1-{M\over
R}\biggr)^2}} + R^2(d\theta^2 + \sin^2\theta\,d\phi^2).\eqno(1.2)
$$
We find that the spacetime metric is static inside as well outside of the horizon
$R=M$ with respect to the time $T$. Could this difference in the dynamical properties be related to
the result that non-extreme holes radiate, whereas extreme do not?

         In this paper we shall address this question by means of a simple
quantum mechanical model of the Reissner-Nordstr\"{o}m black hole. Our model
gives a Hamiltonian quantum theory of spherically symmetric, asymptotically
flat electrovacuum spacetimes. The model is based on the study of the
Hamiltonian dynamics of such spacetimes by Louko and Winters-Hilt in
Ref.$\lbrack 5\rbrack$. In Ref.$\lbrack 5\rbrack$ it was found that after the
classical constraints of the ADM formulation of the dynamics of such spacetimes
have been solved, only two independent degrees of freedom, together with the
corresponding canonical momenta, are left. In other words, the real
physical phase space of spherically symmetric, asymptotically flat
electrovacuum spacetimes is four-dimensional. Because of that, the Hamiltonian
quantum theory of such spacetimes can be addressed by means of a
finite-dimensional quantum mechanics.

          After performing in Section 2 a minor modification of the analysis
made in Ref.$\lbrack 5\rbrack$ at the classical level, we shall in Section 3
construct a Hamiltonian quantum theory of electrovacuum spacetimes under study.
To make things simple, the electric charge is kept fixed. As a consequence, the
phase space contains geometrodynamical variables only. The crucial point is the
choice of phase space coordinates in such a way that they reflect the dynamical
properties of non-extreme Reissner-Nordstr\"{o}m black hole in a natural
manner. To put it simply, we shall use the radius of the wormhole throat of the
Reissner-Nordstr\"{o}m hole, from the point of view of an observer in a radial
free fall through the bifurcation two-sphere, as the configuration variable of
our theory. (For the conformal diagram of maximally extended
Reissner-Nordstr\"{o}m spacetimes see, for example, Refs.$\lbrack 6\rbrack$
and $\lbrack 7\rbrack$.) That radius takes it minimum value $M-\sqrt{M^2-Q^2}$
at the past $R=r_{-}$-hypersurface, then attains its maximum value
$M +\sqrt{M^2-Q^2}$ at the bifurcation two-sphere, and finally shrinks back to
its minimum value $M-\sqrt{M^2-Q^2}$ at the future $R=r_{-}$-hypersurface. As
one can see, the domain of the classically allowed values of the throat radius
includes just one value, $M$, in the limit of extremality. Hence, extreme black
holes have no dynamics with respect to our dynamical variable.

               It turns out that the choice of the momentum variable conjugate
to the throat radius is related to the choice of the foliation of spacetime
into space and time. In this paper we choose a foliation in which the time
coordinate at the right infinity is chosen to be an asymptotic Minkowski time,
and the time evolution at the left infinity is frozen. The time coordinate at the
throat is chosen in such a way that the proper time of an observer in a free
fall through the bifurcation two-sphere agrees with the asymptotic Minkowski
time. With this choice of foliation, the classical Hamiltonian is quadratic in
momenta and its numerical value agrees, when the electric potential is assumed
to vanish at infinity, with the ADM mass of the hole, from the point of view of
a distant observer at rest with respect to the hole. The classical Hamiltonian
is then replaced by the corresponding self-adjoint Hamiltonian operator, and the
spectrum of that operator, which gives the ADM mass spectrum of the hole, is
analyzed.

             The Hamiltonian quantization outlined above was performed for
the Schwarz- schild black hole in Ref.[8], where also the Hamiltonian quantum
theory of Reissner-Nordstr\"{o}m black holes of Section 3 of the present paper
was outlined on a qualitative level. In this paper, however, that Hamiltonian
quantum theory is developed in full details. We shall see that the mass
spectrum is discrete and bounded below. Moreover, we shall see that with an
appropriate choice of a self-adjoint extension, the spectrum of the quantity
$$
M^2-Q^2
$$
is {\it strictly positive}. Regarding Hawking radiation, this is a very
interesting result. If we think of Hawking radiation as an outcome of a chain
of transitions from higher to lower energy eigenstates of the hole, the
positivity of the spectrum of the quantity $M^2-Q^2$ implies that a
non-extreme black hole with non-zero temperature can never become, through
Hawking radiation, an extreme black hole with zero temperature. This result is
in harmony with the third law of black hole thermodynamics as well as with the
qualitative difference between extreme and non-extreme black holes.

           The WKB analysis of the eigenvalue equation for the ADM mass of the
hole yields the result that for macroscopic black holes the eigenvalues of the
quantity $\sqrt{M^2-Q^2}$ are of the form $\sqrt{2n}$, where $n$ is an integer.
The physical interest of this result lies in its relationship with the proposal
made first by Bekenstein in 1974[9]. He argued that the possible eigenvalues of
the black hole horizon area are of the form:
$$
A_n = \gamma nl_{Pl}^2,\eqno(1.3)
$$
where $\gamma$ is a pure number of order one, $n$ is an integer, and
$l_{Pl}:=(\hbar G\slash c^3)^{1\over 2}$ is the Planck length. In Ref.[8] the
spectrum (1.3) was, in effect, obtained for the area of the Schwarzschild black
hole with $\gamma=32\pi$, by means of a model similar to the one used here. In
this paper we shall see that although the spectrum of the area of the outer
horizon of the Reissner-Nordstr\"{o}m black hole fails to coincide with
Eq.(1.3), the eigenvalues of the sum of the areas of the inner and the outer
horizons of the hole
--which we shall refer, for the sake of covenience, as the {\it total} area of
the horizons-- are, in the semi-classical limit, of the form:
$$
A_n^{tot} \approx 32\pi nl_{Pl}^2 + 2A^{ext},\eqno(1.4)
$$
where $A^{ext}:=4\pi Q^2$ is the area of an extreme black hole. Hence, our
result on the spectrum of the total area of the Reissner-Nordstr\"{o}m black
hole is closely related to Bekenstein's proposal: according to Bekenstein's
proposal, the spectrum of the area of the {\it outer horizon} of a black hole is of
the form (1.3), whereas our model implies that it is the {\it total} area of
the hole
which is quantized in that manner.  Moreover, our model implies
that, up to the factor $2A^{ext}$, the spectrum of the total area of the
horizons of the hole is exactly the same for the Reissner-Nordstr\"{o}m and
Schwarzschild black holes.

         Finally, at the end of this paper, we shall relax the requirement that
 the electric charge is a mere external parameter of our theory. Instead, we
shall consider the electric charge as a dynamical variable accounting for the
electromagnetic degrees of freedom of the hole. The resulting quantum theory
implies that the spectrum of the electric charge of the hole is discrete.

\vfill\eject

\centerline{\title 2. Hamiltonian Reduction}

\bigskip

    In the curvature coordinates $R$ and $T$ the spacetime metric corresponding
to the Reissner-Nordstr\"{o}m black hole can be written as:
$$
ds^2 = -\biggl(1-{{2M}\over R} + {{Q^2}\over{R^2}}\biggr)\,dT^2 +
{{dR^2}\over{1-{{2M}\over R} + {{Q^2}\over{R^2}}}} +
R^2\,d\Omega^2.\eqno(2.1)
$$
In this expression, $M$ is the mass and $Q$ is the electric charge of the hole.
$d\Omega^2$ is the line element on the unit two-sphere. As it is well known,
the Reissner-Nordstr\"{o}m metric is the only spherically symmetric,
asymptotically flat electrovacuum solution to the combined
Einstein-Maxwell equations.
In the coordinates $R$ and $T$ the only non-zero component of the
electromagnetic potential $A_\mu$ is
$$
A_T = {Q\over R}.\eqno(2.2)
$$

     Before going into the canonical quantization of Reissner-Nordstr\"{o}m
spacetimes we must investigate the Hamiltonian dynamics of such spacetimes. An
extensive study of this problem, along the lines first shown by Kucha\v{r}[10]
in
the case of Schwarzschild spacetimes, was performed by Louko and Winters-Hilt in
Ref.[5]. However, the considerations of those authors were
thermodynamically motivated, and so they restricted their investigations to the
exterior regions of the hole, whereas our interest, in contrast, lies basically
in the interior regions of the hole. In what follows, we shall briefly review
the analysis of Ref.[5], and consider an extension of that analysis to include
also the interior regions of the hole.

     The starting point in Ref.[5] was to write the general spherically
symmetric Arnowitt-Deser-Misner (ADM) metric in the form:
$$
ds^2 = -N^2\,dt^2 + \Lambda^2(dr + N^r\,dt)^2 + R^2\,d\Omega^2.\eqno(2.3)
$$
In this equation, the lapse $N$ and the shift $N^r$, as well as the variables
$\Lambda$ and $R$, which are considered as the dynamical variables of the
spacetime geometry, are assumed to be functions of the time coordinate $t$ and
the radial coordinate $r$ only. The electromagnetic potential $A_\mu$, also, is
assumed to be spherically symmetric such that its only non-zero components are
$$
\eqalignno{A_t &:= \phi,&(2.4.a)\cr
           A_r &:= \Gamma,&(2.4.b)\cr}
$$
where $\phi$ and $\Gamma$ are assumed to be functions of $t$ and
$r$ only.

      In Ref.[5], the radial coordinate $r$ was taken to be within the interval
$[0,\infty)$. To understand the meaning of this semi-boundedness of the allowed
values of $r$, recall the conformal diagram of the maximally extended
Reissner-Nordstr\"{o}m spacetime (such a diagram can be seen, for example, in
Refs.[6] and [7]). The maximally extended Reissner-Nordstr\"{o}m spacetime has,
in the interior regions of the hole, a periodic geometrical structure. We may
choose one such period and pick up from the conformal diagram a bifurcation
point corresponding to that period. In the bifurcation point, two lines with
the curvature coordinate $R=r_{+}$, where
$$
r_{+} := M + \sqrt{M^2-Q^2}\eqno(2.5)
$$
is the radius of the outer horizon of the hole, intersect each other. In
Ref.[5], the point $r=0$ corresponds to the bifurcation point, and
$r\longrightarrow\infty$ corresponds to the right hand side infinity in the
conformal diagram. Thus, the spacelike hypersurfaces where the time coordinate
$t$ is a constant begin from the bifurcation point and extend to the asymptotic
infinity. Hence, these hypersurfaces can never cross the apparent horizon
$R=r_{+}$, and the investigation performed in Ref.[5] is restricted to the
exterior regions of the hole.

      In this paper, we take the range of $r$ to be from negative to positive
infinity. The region $r\longrightarrow -\infty$ corresponds to the left, and
the region $r\longrightarrow +\infty$ to the right hand side asymptotic
infinity in the conformal diagram. The spacelike hypersurfaces where the time
$t$ is a constant, are assumed to go through the interior regions of the hole
in arbitrary ways. However, the requirement that the hypersurfaces $t=constant$
are spacelike and extendible from left to right asymptotic infinities imposes
an important restriction: if we look at the conformal diagram of the
Reissner-Nordstr\"{o}m spacetime, we find that it is not possible to push
these hypersurfaces beyond the inner horizons, where
$$
R = r_{-} := M-\sqrt{M^2-Q^2},\eqno(2.6)
$$
since otherwise the hypersurfaces would necessarily fail to be spacelike.
Hence, our study of the Hamiltonian dynamics of Reissner-Nordstr\"{o}m
spacetimes must be restricted to include, in addition to the left and right
exterior regions of the hole, only such an interior region of the hole which
lies between two successive $R=r_{-}$-hypersurfaces in the conformal diagram.
Our spacelike hypersurface $t=constant$ therefore begins its life at the past
$R=r_{-}$-hypersurface, then goes through the bifurcation point $R=r_{+}$, and
finally ends its life at the future $R=r_{-}$-hypersurface. Bearing this
restriction in mind, we shall now go into the Hamiltonian dynamics of
Reissner-Nordstr\"{o}m spacetimes. In most technical details, the discussion
goes exactly as in Ref.[5]. The major difference comes from the boundary
terms.

      To begin with, one writes the action. For the Einstein-Maxwell theory the
action is, in general,
$$
S = {1\over{16\pi}}\int d^4x\,\sqrt{-g} (^{(4)}R - F_{\mu\nu}F^{\mu\nu}) + (boundary
\,\,\,terms).\eqno(2.7)
$$
In this equation, the integration  is performed over the whole spacetime under
consideration, $^{(4)}R$ is the four-dimensional Riemann scalar of spacetime,
and
$$
F_{\mu\nu} := \partial_\mu A_\nu - \partial_\nu A_\mu\eqno(2.8)
$$
is the electromagnetic field tensor. As it was shown in Ref.[5], the action
takes, when the ans\"{a}tze (2.3) and (2.4) are used, the form:
$$
S = S_\Sigma + S_{\partial\Sigma},\eqno(2.9)
$$
where
$$
S_\Sigma := \int dt\int_{-\infty}^{+\infty}dr\,(P_R\dot{R} + P_\Lambda\dot{\Lambda} +
P_\Gamma\dot{\Gamma} - N{\cal H} - N^r{\cal H}_r - {\tilde\phi}G), \eqno(2.10)
$$
and $S_{\partial\Sigma}$ is a boundary term corresponding to the left and right
hand side asymptotic infinities. In Eq.(2.10), the overdot means time
derivative, and $P_R$, $P_\Lambda$, $P_\Gamma$, respectively, are the canonical
momenta conjugate to the variables $R$, $\Lambda$ and $\Gamma$. $\cal{H}$,
${\cal H}_r$ and $G$, respectively, are the Hamiltonian, diffeomorphism and
Gaussian constraints. For the explicit expressions for the momenta $P_R$ and
$P_\Lambda$ as well as the constraints ${\cal H}$ and ${\cal H}_r$ we refer the
reader to the Ref.[5]. Of particular interest is the momentum $P_\Gamma$. It
can be written as:
$$
P_\Gamma = N^{-1}\Lambda^{-1}R^2({\dot \Gamma} - \phi'),\eqno(2.11)
$$
where $\phi':={\partial\over{\partial r}}\phi$. The Gaussian constraint is
$$
G=-P'_\Gamma,\eqno(2.12)
$$
and the quantity ${\tilde \phi}$, which appears as a Lagrangian multiplier of
the theory, is defined as
$$
{\tilde \phi} := \phi - N^r\Gamma.\eqno(2.13)
$$
As to the momenta $P_R$ and $P_\Lambda$, and the constraints ${\cal H}$ and
${\cal H}_r$, it should be noted that since the analysis made in Ref.[5] was
thermodynamically motivated, the authors of that paper included a negative
cosmological constant to improve the convergence of certain integrals. In the
absence of the cosmological constant, however, the expressions for the momenta
and the constraints can be obtained from those in Ref.[5] simply by putting the
cosmological constant to zero.

        What about the boundary term $S_{\partial\Sigma}$ corresponding to the
asymptotic spacelike infinities? To find an expression to the boundary term one
must first specify the fall-off conditions for the canonical variables and the
Lagrangian multipliers of the theory. We shall adopt the same asymptotic
boundary conditions as in Ref.[5] in the absence of a cosmological constant:
$$
\eqalignno{\Lambda(t,r) &= 1 +
 M_\pm(t)\vert r\vert^ {-1} + O^\infty(\vert
r\vert^{-1-\epsilon}),&(2.14.a)\cr
           R(t,r) &= \vert r\vert + O^\infty(\vert
r\vert^{-\epsilon},&(2.14.b)\cr
           P_\Lambda(t,r) &= O(\vert r\vert^{-\epsilon}),&(2.14.c)\cr
           P_R(t,r) &= O^\infty(\vert
r\vert^{-1-\epsilon}),&(2.14.d)\cr
           N(t,r) &= N_\pm(t) + O^\infty(\vert r\vert^{-\epsilon}),&(2.14.e)\cr
           N^r(t,r) &= O^\infty(\vert r\vert^{-\epsilon}),&(2.14.f)\cr
           \Gamma(t,r) &= O^\infty(\vert r\vert^{-1-\epsilon}),&(2.14.g)\cr
           P_\Gamma(t,r) &= Q_\pm(t) + O^\infty(\vert
r\vert^{-\epsilon}),&(2.14.h)\cr
           {\tilde \phi}(t,r) &= {\tilde \phi}_\pm(t) + O^\infty(\vert
r\vert^{-\epsilon}),&(2.14.i)\cr}
$$
when $r\longrightarrow\pm\infty$. In these equations, $\epsilon > 0$,
and $O(\vert r\vert^{-\epsilon})$ denotes a term that falls off at infinity
as $\vert r\vert^{-\epsilon}$ and whose derivatives with respect to $\vert
r\vert$ fall accordingly as $\vert r\vert^{-\epsilon-k}$, where $k=1,2,3,...$.
Our fall-off conditions ensure that spacetime is asymptotically flat at
asymptotic infinities. Of particular interest is the fall-off condition
(2.14.f), which states that the shift vanishes at infinities. This means that
our asymptotic coordinate systems are at rest with respect to the hole.

        The boundary term $S_{\partial \Sigma}$ has now two parts. The first of
them is:
$$
S_{\partial\Sigma}^{ADM} := -\int dt\,(N_+(t)E_+(t) + N_-(t)E_-(t)),\eqno(2.15)
$$
where
$$
E_\pm(t) := {1\over{16\pi}}\int_{S_\pm^2}
dS^{a}\delta^{bc}(g_{ab,c}-g_{bc,a}),\eqno(2.16)
$$
are the ADM energies at the right and left infinities. In Eq.(2.16), the
integral is taken over two-spheres $\vert r\vert =constant$ at infinities. The
term (2.15) is needed in order to cancel the terms appearing at spatial
infinities when the action $S_\Sigma$ is varied with respect to the the
variables $R$ and $\Lambda$ and their canonical momenta. One can see from
Eqs.(2.14.a) and (2.14.b) that the boundary term  $S_{\partial\Sigma}^{ADM}$ in
Eq.(2.15) takes the form:
$$
S_{\partial\Sigma}^{ADM} = -\int dt\,(N_+(t)M_+(t) + N_-(t)M_-(t)). \eqno(2.17)
$$

     The second part of the boundary term is related to electromagnetism. If
one writes down all of the electromagnetic terms in the action
$S_\Sigma$ of Eq.(2.10), one finds that the only term involving spatial
derivatives of the quantities $\Gamma$, $\phi$ and $P_\Gamma$ is:
$$
\int dt \int_{-\infty}^{+\infty} dr\, P'_\Gamma\phi.
$$
Varying the action $S_\Sigma$ with respect to $P_\Gamma$ therefore brings along
a term
$$
\int dt \int_{-\infty}^{+\infty} dr\,(\delta P_\Gamma\phi)'.
$$
Unless the variation of $P_\Gamma$ is assumed to vanish at spatial infinities,
we are thus compelled to bring along a boundary term
$$
S_{\partial\Sigma}^{em} = -\int dt\,(Q_+(t){\tilde\phi}_+(t) -
Q_-(t){\tilde\phi}_-(t)).\eqno(2.18)
$$
The whole boundary term $S_{\partial\Sigma}$ is the sum the terms
$S_{\partial\Sigma}^{ADM}$ and $S_{\partial\Sigma}^{em}$, and we find that the
whole action finally becomes to:
$$
\eqalign{S &= \int dt\,\int_{-\infty}^{+\infty} dr\,(P_R{\dot R}
            + P_\Lambda{\dot\Lambda}
+ P_\Gamma{\dot\Gamma} - N{\cal H} - N^r{\cal H}_r - {\tilde\phi}G)\cr
           &-\int dt\,(N_+ M_+ + N_-M_- + {\tilde\phi}_+Q_+ -
{\tilde\phi}_-Q_-).\cr}\eqno(2.19)
$$

     As it was mentioned before, the Reissner-Nordstr\"{o}m solution is the
only spherically symmetric, asymptotically flat solution to Einstein-Maxwell
equations. This solution, moreover, is completely characterized by the mass and
the charge parameters $M$ and $Q$. It was found in Ref.[5] that one can read
off these parameters from any small piece of spacetime if one knows the values
of the phase space coordinates $\Lambda$, $R$, $\Gamma$, $P_\Lambda$, $P_R$ and
$P_\Gamma$ in that region. More precisely,
$$
\eqalignno{M &= {1\over 2}{{P_\Lambda^2}\over R} + {1\over 2}{{P_\Gamma^2}\over
R} + {1\over 2}R - {1\over 2}{{R(R')^2}\over{\Lambda^2}},&(2.20.a)\cr
           Q &= P_\Gamma.&(2.20.b)\cr}
$$
Although the derivation of these equations in Ref.[5] was performed in the
exterior region of the hole, the same arguments as in Ref.[5] go through in any
region of spacetime. In curvature coordinates, the reader may check the
validity of Eq.(2.20.b) by using Eqs.(2.2), (2.4) and (2.11).

  In addition to the parameters $M$ and $Q$, ane can also read off how the
spacelike hypersurface has been embedded into the Reissner-Nordstr\"{o}m
spacetime from the values of the phase space coordinates in any point of the
hypersurface. In Ref.[5] it was found that if the curvature coordinate $T$ is
considered as an arbitrary function of the coordinates $t$ and $r$, then
$$
-T' = R^{-1}F^{-1}\Lambda P_\Lambda P_\Gamma,\eqno(2.21)
$$
where
$$
F:= \biggl({{R'}\over\Lambda}\biggr)^2 - \biggl({{P_\Lambda}\over
R}\biggr)^2.\eqno(2.22)
$$
>From Eq.(2.21) one can solve $T$ as a function of $r$ provided that one knows
$T$ for one value of $r$. Keeping $t$ as a constant one can read off the
position of the $t=constant$ hypersurface of spacetime in the
Reissner-Nordstr\"{o}m manifold from that solution.

    The solution (2.2) to Maxwell's equations involves a specific fixing of the
electromagnetic gauge. The general spherically symmetric solution can be
written as
$$
\eqalignno{A_T &= {Q\over R} + {\partial\over{\partial T}}\xi,&(2.23.a)\cr
           A_R &= {\partial\over{\partial R}}\xi,&(2.23.b)\cr}
$$
where $\xi$ is an arbitrary function of $T$ and $R$. Hence, the general
solution in the coordinates $t$ and $r$ is
$$
\eqalignno{A_t &= {Q\over R}{\dot T} + {\dot\xi},&(2.24.a)\cr
           A_r &= {Q\over R}T' + \xi'.&(2.24.b)\cr}
$$
Comparing Eqs.(2.4.b), (2.20.b), (2.21) and (2.24.b)  one finds that:
$$
\xi' = \Gamma + R^{-2}F^{-1}\Lambda P_\Lambda P_\Gamma.\eqno(2.25)
$$
If one knows the gauge function $\xi$ for one value of $r$, one can solve $\xi$
for any $r$ from Eq.(2.25). Hence, information about the choice of the
electromagnetic gauge is carried by the phase space coordinates of the
theory.

     At first sight, there might seem to be a difficulty with the
horizons $R=r_\pm$, because it follows from Eqs.(2.20) and (2.22) that
$$
F = 1 - {{2M}\over R} + {{Q^2}\over{R^2}}.\eqno(2.26)
$$
Hence, the quantities $T'$ and $\xi'$ appear to have a singularity when
$R=r_\pm$. This problem was considered in the case $Q=0$ by Kucha\v{r}[10]. His
conclusion was that one can nevertheless propagate through horizon one's
knowledge of $T$ by using Eq.(2.21). The very same arguments can be applied
also when $Q\neq 0$. Hence, the event horizon does not pose a problem for
our knowledge of $T$ and $\xi$ as functions of $r$.

     In Ref.[5], the idea was to "forget" for a moment the fact that $M$ and
$Q$ are the mass and the charge parameters of the Reissner-Nordstr\"{o}m black
hole, and instead consider $M$ and $Q$ as arbitrary functions of $r$ and $t$,
whose relationship with the phase space coordinates $R$, $\Lambda$, $\Gamma$,
$P_R$, $P_\Lambda$ and $P_\Gamma$ is given by Eq.(2.20). A canonical
transformation was then performed from the original phase space coordinates
to the new canonical variables such that $M$ and $Q$ were considered as the
configuration variables, and the quantities $-T'$ and $-\xi'$ defined in terms
of the original phase space
coordinates as in Eqs.(2.21) and (2.25) were taken to be the corresponding
canonical momenta $P_M$ and $P_Q$. A new canonical momentum $P_{\R}$
conjugate to $R$ was also defined such that the whole transformation from the
"old" to the "new" phase space variables was canonical.

           What about the constraints? Varying the action (2.19) with respect
to the functions $N$, $N^r$ and ${\tilde\phi}$ yields the following constraint
equations:
$$
\eqalignno{{\cal H} &= 0,&(2.27.a)\cr
           {\cal H}_r &= 0,&(2.27.b)\cr
              G &= 0.&(2.27.c)\cr}
$$
It should be noted that one is not allowed to vary the action with respect
to the functions $N_\pm$ and ${\tilde\phi}_\pm$ but these functions should be
kept as prescribed functions of the time $t$. That is because varying the
action with respect to $N_\pm$ would imply vanishing ADM mass --and hence
flatness-- of spacetime; varying action with respect to ${\tilde \phi}_\pm$, in
turn, would imply vanishing electric charge.

        Using Eqs.(2.12) and (2.20), and the expressions of Ref.[5] for the
Hamiltonian and the diffeomorphism constraints ${\cal H}$ and ${\cal H}_r$ in
the absence of the cosmological constant, one finds that the spatial derivatives
of the mass and the charge functions $M$ and $Q$ can be written in terms of the
constraints:
$$
\eqalignno{M' &= -\Lambda^{-1}R'{\cal H} - \Lambda^{-1}R^{-1}P_\Lambda{\cal
H}_r + (\Lambda^{-1}R^{-1}\Gamma P_\Gamma - R^{-1}P_\Gamma)G,&(2.28.a)\cr
           Q' &= -G.&(2.28.b)\cr}
$$
Hence, the constraint equations (2.27) imply that:
$$
\eqalignno{M' &= 0,&(2.29.a)\cr
           Q' &= 0.&(2.29.b)\cr}
$$
In other words, the mass and charge functions $M$ and $Q$ are constants with
respect to $r$. Moreover, it was noted in Ref.[5] that
$$
{\cal H}_r = P_MM' + P_QQ' + P_{\R}R'.\eqno(2.30)
$$
Eqs.(2.27) and (2.29) therefore imply:
$$
P_{\R} = 0.\eqno(2.31)
$$
Hence, the action written in terms of the variables $M$, $Q$, $R$ and their
canonical momenta,
$$
\eqalign{S &= \int dt\int_{-\infty}^{+\infty} dr\,(P_M{\dot M} + P_Q{\dot Q} +
P_\R{\dot R} - N{\cal H} - N^r{\cal H}_r - {\tilde\phi}G)\cr
           &- \int dt\,(N_+M_+ + N_-M_- + {\tilde\phi}_+Q_+ -
{\tilde\phi}_-Q_-),\cr}\eqno(2.32)
$$
takes, when the constraint equations (2.27) are satisfied, the form
$$
S = \int dt\lbrack p_m{\dot m} + p_q{\dot q} - (N_+ + N_-)m - ({\tilde\phi}_+ -
{\tilde\phi}_-)q\rbrack,\eqno(2.33)
$$
where we have defined:
$$
\eqalignno{m(t) &:= M(t,r),&(2.34.a)\cr
           q(t) &:= Q(t,r),&(2.34.b)\cr
         p_m(t) &:= \int_{-\infty}^{+\infty} dr\,P_M(t,r),&(2.34.c)\cr
         p_q(t) &:= \int_{-\infty}^{+\infty} dr\,P_Q(t,r).&(2.34.d)\cr}
$$
When the constraint equations are satisfied, our infinite-dimensional
phase space is thus reduced to
a phase space which is spanned by just four canonical coordinates. These
canonical coordinates are the variables $m$ and $q$ --which can be identified as
the mass $M$ and the charge $Q$ of the hole when Einstein's equations are
satisfied-- and the corresponding canonical momenta $p_m$ and $p_q$.

       The momenta $p_m$ and $p_q$ have an interesting interpretation: because
we have defined $P_M:=-T'$ and $P_Q:=-\xi'$, we find that $p_m$ is simply the
difference in the Minkowski time $T$ at the left and right infinities on the
spacelike hypersurface $t=constant$. The momentum $p_q$, in turn, tells the
difference between the choices of the gauge function $\xi$ at the
asymptotic infinities of spacetime.

       One can read off from Eq.(2.33) the true reduced Hamiltonian of the
Reissner-Nordstr\"{o}m hole in terms of the variables $m$ and $q$:
$$
H = (N_+ + N_-)m + ({\tilde\phi}_+ - {\tilde\phi}_-)q.\eqno(2.35)
$$
The Hamiltonian equations of motion are therefore
$$
\eqalignno{{\dot m} &= {{\partial H}\over{\partial p_m}} = 0,&(2.36.a)\cr
           {\dot q} &= {{\partial H}\over{\partial p_q}} = 0,&(2.36.b)\cr
         {\dot p}_m &= -{{\partial H}\over{\partial m}} = -(N_+ +
N_-),&(2.36.c)\cr
         {\dot p}_q &= -{{\partial H}\over{\partial q}} = -({\tilde\phi}_+ -
{\tilde\phi}_-).&(2.36.d)\cr}
$$
As one can see from Eqs.(2.36.a) and (2.36.b), the mass $m$ and the charge $q$
are constants of motion of the system.

\bigskip

\bigskip

\centerline{\title 3. Quantum Theory with Charge as an}
\centerline{\title External Parameter}

\bigskip

      After finding in Eq.(2.35) an expression to the classical reduced
Hamiltonian of Reissner-Nordstr\"{o}m spacetimes in terms of the variables $m$
and $q$ and the corresponding canonical momenta, we shall now proceed into a
Hamiltonian quantization of such spacetimes.  The aim of all physical theories,
at least in principle, is to be able to predict
the possible outcomes of measurements. When we talk about measurements,
however, we need a reference to an {\it observer} performing these
measurements: the possible outcomes of measurements are the possible outcomes
of measurements as such as they can be measured, in principle, by a certain
observer. The properties of the observer, in turn, motivate the structure of
the theory.\footnote{$^1$}{Recently, an interesting point of view to the
interpretation of quantum mechanics was suggested by Rovelli[11]. His idea was,
in rough terms, that one is not justified to talk about any {\it absolute}
quantum state of a physical system. Instead, one should talk about a quantum
state relative to some observer. This idea has given some inspiration to
the point of view adopted in this paper.}

       In this paper we choose the observer in a most simple manner: our
observer is at the right hand side asymptotic infinity in the conformal
diagram, at rest with respect to
the Reissner-Nordstr\"{o}m hole.  Our aim is to construct a quantum theory
of the Reissner-Nordstr\"{o}m spacetime from the point of view of such an
observer.  To this end, we choose the lapse
functions $N_\pm(t)$ at asymptotic infinities as:
$$
\eqalignno{N_+(t) &\equiv 1,&(3.1.a)\cr
           N_-(t) &\equiv 0.&(3.1.b)\cr}
$$
In other words, we have chosen the time coordinate at the right infinity to be
the proper time of our observer, and we have "frozen" the time evolution at the
left infinity. This can be considered justified on grounds that our observer
can make observations at just one infinity.

      The next task is to fix the functions ${\tilde\phi}_\pm(t)$. As one can
see from Eqs.(2.13.) and (2.14), the functions ${\tilde\phi}_\pm(t)$ are just
the electric potentials at asymptotic infinities. It is customary to choose the
zero point of the electric potential in such a way that at asymptotic
infinities the electric potential vanishes. As one can see from Eq.(2.2), this
choice is compatible with the Reissner-Nordstr\"{o}m solution to
Einstein-Maxwell equations. Hence, we choose:
$$
{\tilde\phi}_+(t) \equiv {\tilde\phi}_-(t) \equiv 0.\eqno(3.2)
$$
With these choices of the lapse functions and the electric potentials the
reduced Hamiltonian (2.35) takes the form:
$$
H = m.\eqno(3.3)
$$
Because of that, the numerical value of our Hamiltonian is just the mass $M$ of
the hole. This mass includes, from the point of view of our observer, all the
energy of the system, gravitational as well as electromagnetic.

      Now, one could, of course, use the variables $m$ and $q$ and their
canonical momenta as the phase space coordinates of the system, and construct a
Hamiltonian quantum theory of the Reissner-Nordstr\"{o}m hole based on the use
of these coordinates. There is, however, a very grave disadvantage with these
coordinates: They describe the {\it static} aspects of the black hole spacetime
only. Indeed, we saw in Eq.(2.36) that the variables $m$ and $q$ are constants
of motion of the system. However, there is {\it dynamics} in the
Reissner-Nordstr\"{o}m spacetime in the sense that in the region where
$r_-<R<r_+$, there is no timelike Killing vector field orthogonal to a
spacelike hypersurface. Our task is to find such
phase space coordinates which describe the dynamics of spacetime in a natural
manner.

     When choosing the phase space coordinates we again refer to the properties
of our observer. Our observer sees the exterior regions of the black hole as
static, and he is an inertial observer. These properties prompt us to choose
the phase space coordinates in such a manner that when the classical equations
of motion are satisfied, all the dynamics is, in a certain sense, confined
inside the apparent horizon $R=r_+$ of the hole. Moreover, as we shall see in a
moment, the choice of the phase space coordinates describing the dynamics of
spacetime is related to the choice of slicing of spacetime into space and
time. We choose a slicing where the proper time of an observer in a radial
free fall through the bifurcation two-sphere coincides with the proper time of
our faraway observer at rest. On grounds of the Principle of Equivalence
one may view these kind of slicings to be in a preferred position in relating
the physical properties of the black hole interior to the
physics observed by our faraway observer.

    The new phase space coordinates describing the dynamics of the black hole
spacetime can be obtained from the phase space coordinates $m$, $q$, $p_m$ and
$p_q$ by means of an appropriate canonical transformation. To make things
simple, we shall in this Section consider the charge $q$ as a mere external
parameter of the system, having a fixed value $Q$. In the next Section, the
charge will be considered as a dynamical variable. Hence, the dimension of the
phase space of our system is, in this Section, just two.

     As in Ref.[8], we now perform a canonical transformation from the phase
space variables $(m,p_m)$ to the new phase pace variables $(a,p_a)$ such that
the relationship between the "old" and the "new" phase space variables is ($q$
is now a constant, which we denote by $Q$):
$$
\eqalignno{\vert p_m\vert &= \sqrt{2ma-a^2-Q^2} +
m\sin^{-1}\biggl({{m-a}\over{\sqrt{m^2-Q^2}}}\biggr) + {1\over 2}\pi
m,&(3.4.a)\cr
                      p_a &= sgn(p_m)\sqrt{2ma - a^2 - Q^2},&(3.4.b)\cr}
$$
and we have imposed by hand a restriction:
$$
-\pi m \leq p_m \leq \pi m.\eqno(3.5)
$$
With the restriction (3.5) the transformation (3.4) is well-defined and
one-to-one. It follows from Eq.(3.4.b) that
$$
m = {{p_a^2}\over{2a}} + {1\over 2}a + {{Q^2}\over{2a}}.\eqno(3.6)
$$
If one substitutes this expression for $m$ to Eq.(3.4.a), one gets $p_m$ in
terms of $a$ and $p_a$. One finds that the fundamental Poisson brackets between
$m$ and $p_m$ are preserved invariant, and hence the transformation (3.4) is
canonical.

        Eqs.(3.3) and (3.6) imply that the classical Hamiltonian takes, in
terms of the variables $a$ and $p_a$, a form:
$$
H = {{p_a^2}\over{2a}} + {1\over 2}a + {{Q^2}\over{2a}}.\eqno(3.7)
$$
The geometrical interpretation of the variable $a$ is extremely easy to find.
We first write the Hamiltonian equation of motion for $a$:
$$
{\dot a} = {{\partial H}\over{\partial p_a}} = {{p_a}\over a},\eqno(3.8)
$$
and it follows from Eq.(3.6) and the fact that $m=M$ when the classical
equations of motion are satisfied that the equation of motion for $a$ is
$$
{\dot a}^2 = {{2M}\over a} - 1 -{{Q^2}\over{a^2}}.\eqno(3.9)
$$
One can see from the Reissner-Nordstr\"{o}m metric in Eq.(2.1) that the
equation of motion for an observer in a radial free fall through the
bifurcation two-sphere is
$$
{\dot R}^2 = {{2M}\over R} - 1 - {{Q^2}\over{R^2}},\eqno(3.10)
$$
where the overdot means proper time derivative. As one can see, Eqs.(3.9) and
(3.10) are identical. Hence, we can interpret $a$ as the radius of the wormhole
throat of the Reissner-Nordstr\"{o}m black hole, from the point of view of an
observer in a radial free fall through the bifurcation two-sphere. Moreover, we
see from Eq.(3.9) that $a$ is confined to be, classically, within the region
$\lbrack r_-,r_+\rbrack$. In other words, our variable $a$ "lives" only within
the inner and outer horizons of the Reissner-Nordstr\"{o}m black hole, and this
is precisely the region in which it is impossible to find a time coordinate in
such a way that spacetime with respect to that time coordinate would be static.
Hence, both of the requirements we posed for our phase space coordinates are
satisfied: dynamics is confined inside the apparent horizon and the time
coordinate on the wormhole throat is the proper time of an observer in a
radial free fall.

    With the interpretation explained above, the restriction (3.5) becomes
understandable. One can see from Eq.(2.36.c) that when the lapse functions
$N_\pm$ at asymptotic infinities are chosen as in Eq.(3.1), the canonical
momentum $p_m$ conjugate to $m$ is $-t + constant$, where $t$ is the time
coordinate of our asymptotic observer. Now, the transformation (3.4) involves
an identification of the time coordinate $t$ with the proper time of a freely
falling observer on the throat. However, as it was noted in the beginning of
Section 2, it is impossible to to push the spacelike hypersurfaces $t=constant$
beyond the $R=r_-$ hypersurfaces in the conformal diagram. The proper time a
freely falling observer needs to fall from the past $R=r_-$ hypersurface to the
future $R=r_-$ hypersurface through the bifurcation two-sphere is, as it can
be seen from Eq.(3.10),
$$
\Delta t = 2\int_{r_-}^{r_+} {{R'\,dR'}\over{\sqrt{2MR' - {R'}^2 - Q^2}}} =
2\pi M,\eqno(3.11)
$$
and hence the restriction (3.5) is needed. As one can see from Eq.(3.4.a),
$\vert p_m\vert =0$, when $a=r_+$,
and $\vert p_m\vert = \pi M$, when $a=r_-$. We have chosen $p_m$ to be
posititive, when the hypersurface $t=constant$ lies between the past
$R=r_-$ hypersurface and the bifurcation point, and negative when that
hypersurface lies between the bifurcation point and the future
$R=r_-$ hypersurface.

      As to the classical Hamiltonian theory, the only thing one still needs to
check is, whether there exist such foliations of the Reissner-Nordstr\"{o}m
spacetime where the Minkowski time $t$ at asymptotic infinity and the
proper time of a freely falling observer at the throat through the bifurcation
two-sphere really are the one and the
same time coordinate. It is easy to see that time coordinates determining this
sort of foliations really exist. As a concrete example, consider a
generalization of the so called Novikov coordinates in the Schwarzschild
geometry[7]. More precisely, one takes a collection of freely falling test
particles whose initial three-velocity with respect to the curvature
coordinates is zero when $T=0$. If one relates the radial coordinate of
spacetime to the positions of these test particles when $T=0$, and takes the
time coordinate to be at every spacetime point the proper time of the test
particle falling through that point, one finds that the time coordinate of our
distant observer at rest, the Minkowski time, and the proper time of a freely
falling observer in the throat are the one and the same time coordinate. It
should be noted, however, that all foliations in which the proper time in the
throat and the asymptotic Minkowski time are identified, are incomplete, since
such foliations, in addition of failing to cover regions outside the past and
future $R=r_-$ hypersurfaces, also fail to cover the whole exterior regions of
the hole. More precisely, these foliations are valid only when $-\pi M\leq
t\leq\pi M$.

       After finding a classical Hamiltonian reflecting the dynamical
properties of Reissner-Nordstr\"{o}m spacetimes, we are now prepared to go into
the Hamiltonian quantization of such spacetimes. In what follows, we shall
specify to a particular class of Hamiltonian quantum theories. More precisely,
we choose our Hilbert space to be the space $L^2(\Re^+,a^sda)$ with the inner
product
$$
\langle\psi_1\vert\psi_2\rangle := \int_0^{\infty}
\psi^*_1(a)\psi_2(a)a^s\,da,\eqno(3.12)
$$
where $s$ is some real number. Through the substitution $p_a\longrightarrow
-i{d\slash da}$ we replace the classical Hamiltonian $H$ of Eq.(3.7) with the
corresponding symmetric Hamiltonian operator
$$
{\hat H} := -{1\over 2}a^{-s}{d\over{da}}\biggl(a^{s-1}{d\over{da}}\biggr)
+ {1\over 2}a +
{{Q^2}\over{2a}}.\eqno(3.13)
$$
Since the numerical value of the classical Hamiltonian $H$ is the total (ADM)
energy of the Reissner-Nordstr\"{o}m hole, we can view the eigenvalue equation
$$
{\hat H}\psi(a) = E\psi(a)\eqno(3.14)
$$
 as an eigenvalue equation for the total energy of the hole, from the point of
view of a distant observer at rest.

        Before going into the detailed analysis of Eq.(3.14), let us pause for
a moment to investigate some qualitative aspects of that equation. One finds,
by substituting $M$ for $E$ that Eq.(3.14) can be written in the form:
$$
a^{-s}{d\over{da}}\biggl(a^{s-1}{d\over{da}}\biggr)\psi(a) = \biggl({{Q^2}\over
a} + a - 2M\biggr)\psi(a).\eqno(3.15)
$$
As one can see, the function $({{Q^2}\over a} + a - 2M)$ is negative, when
$r_-<a<r_+$, and positive (or zero) elsewhere. Semiclassically, one may
therefore expect oscillating behaviour from the wave function $\psi(a)$ when
$r_-<a<r_+$, and
exponential behaviour elsewhere. Hence, our system is somewhat analogous to a
particle in a potential well such that $a$ is confined, classically, between the
outer and the inner horizons of the black hole. What happens semiclassically is
that the wave packet corresponding to the variable $a$ is reflected from the
future inner horizon. As a result we get, when the hole is in a stationary
state, a standing wave between the outer and inner horizons. Thus, the classical
incompleteness, associated with the fact that our foliation is valid only when
$-\pi M\leq t\leq \pi M$, is removed by quantum mechanics: in a stationary state
there are no propagating wave packets between the horizons, and our quantum
theory is therefore valid in any moment of time.

       Let us now go into the detailed analysis of the eigenvalue equation
(3.14). To begin with, we see, as in Ref.[8], that if we denote
$$
\eqalignno{x &:= a^{3\slash 2},&(3.16.a)\cr
        \psi &:= x^{-r}\chi(x),&(3.16.b)\cr}
$$
where we have defined
$$
\eqalignno{r &:={{2s-1}\over 6};\,\,\,s\geq 2,&(3.17.a)\cr
           r &:={{7-2s}\over 6};\,\,\,s<2,&(3.17.b)\cr}
$$
then Eq.(3.14) takes the form:
$$
{9\over 8}\biggl[-{{d^2}\over{dx^2}} + {{r(r-1)}\over{x^2}} + {4\over
9}\biggl(x^{2\slash 3} + {{Q^2}\over{x^{2\slash 3}}}\biggr)\biggr]\chi(x) =
E\chi(x).\eqno(3.18)
$$
The Hilbert space is now $L^2(\Re^+,dx)$ with the inner product
$$
\langle\chi_1\vert\chi_2\rangle := \int_0^\infty
\chi^*_1(x)\chi_2(x)\,dx.\eqno(3.19)
$$

     It was shown in Ref.[8] that the energy spectrum in Eq.(3.18) is discrete,
bounded below, and can be made positive. From the physical point of
view, the semi-boundedness and positivity (in some cases) of the spectrum are
very satisfying results: the semi-boundedness of the spectrum implies that one cannot
extract an infinite amount of energy from the system, whereas the positivity of
the spectrum is in harmony with the well-known positive energy theorems of
general relativity which state, roughly speaking, that the ADM energy of
spacetime is
always positive or zero when Einstein's equations are satisfied.

     However, one can prove even more than that. As it is shown in Appendix B,
the eigenvalue equation (3.18) implies that when $r\geq{3\slash 2}$, the
eigenvalues of the quantity
$$
E^2 - Q^2
$$
are {\it strictly positive}, and when ${1\slash 2}\leq r<{3\slash 2}$, the
eigenvalues of the quantity $E^2-Q^2$ can, again, be made positive by means of
an appropriate choice of the boundary conditions of the wave function $\chi(x)$
at the point $x=0$ or, more precisely, by means of an appropriate choice of
a self-adjoint extension. Moreover, the WKB analysis of Eq.(3.18) performed in
Appendix A yields the result that when $\vert Q\vert \gg 1$ and $E^2-Q^2\gg 1$,
such that $r_{-}\geq 1$, the WKB eigenenergies $E_n$ have a property
$$
E_n^2 - Q^2 \sim 2n + 1 + o(1),\eqno(3.20)
$$
where $n$ is an integer and $o(1)$ denotes a term that vanishes asymptotically
for large $n$. We have tested the accuracy of this WKB estimate numerically,
and we have found that, up to the term 1 on the right hand side of Eq.(3.20),
the WKB estimate (3.20) gives fairly accurate results even when $\vert Q\vert$
and $n$ are relatively small (i.e. of order ten).\footnote{$^1$}{We thank Matias
Aunola for performing this numerical analysis to us.} In other words, it seems that
the eigenvalues of the quantity $\sqrt{E^2-Q^2}$ are of the form $\sqrt{2n}$
in the semiclassical limit.

      Now, how should we understand these results? As it was noted in the
Introduction, the positivity of the spectrum of the quantity $E^2-Q^2$ has an
interesting consequence regarding Hawking radiation: if one thinks of Hawking
radiation as a an outcome of a chain of transitions from higher to lower energy
eigenstates, the positivity of the spectrum of $E^2-Q^2$ implies that a
non-extreme Reissner-Nordstr\"{o}m black hole with non-zero temperature can
never become, through Hawking radiation, an extreme black hole with zero
temperature --a result which is in harmony with both the third law of
thermodynamics and the qualitative difference between extreme and non-extreme
black holes. One may consider this result as a strong argument in favour of our
choice of the phase space coordinates describing the dynamics of
Reissner-Nordstr\"{o}m spacetimes.

       How about Eq.(3.20)? Since the area of the apparent horizon of the
Reissner-Nordstr\"{o}m hole is, in natural units,
$$
A_+ := 4\pi(M+\sqrt{M^2-Q^2})^2,\eqno(3.21)
$$
we find that, if $\vert Q\vert$ is negligible when compared to $M$, Eq.(3.20)
yields the result that the possible eigenvalues of the area $A_+$ are, in the
semiclassical limit,
$$
A_{+n} \approx 32\pi nl_{Pl}^2 + constant,\eqno(3.22)
$$
where $l_{Pl}^2:=(\hbar G\slash c^3)^{1\slash 2}$ is the Planck length. This
result is the same as the one obtained in Ref.[8] for a Schwarzschild black hole.
Moreover, the result is in harmony with the proposal (1.3), first made by
Bekenstein in 1974[9] and since then revived by many authors[12-32] on the discrete
area spectrum of a black hole.

       However, if we look at the expression (3.21) for the horizon area, we
find that Eq.(3.20) does {\it not} produce the proposal (1.3) in the general
case. The solution of this dilemma lies in the fact that the
Reissner-Nordstr\"{o}m black hole has, actually, {\it two} horizons: the inner
and the outer horizons. The sum of of the areas of these two horizons --which
we shall call, for the sake of convenience, the {\it total} area of the
horizons--
is
$$
A^{tot} := A_+ + A_-,\eqno(3.23)
$$
where $A_+$ is the area of the outer horizon as given in Eq.(3.21), and
$$
A_- := 4\pi(M-\sqrt{M^2 - Q^2})^2\eqno(3.24)
$$
is the area of the inner horizon. We get
$$
A^{tot} = 16\pi(M^2-Q^2) + 2A^{ext},\eqno(3.25)
$$
where
$$
A^{ext} := 4\pi Q^2\eqno(3.26)
$$
is the area of an extreme black hole. Eq.(3.20) now gives the spectrum of the
total area of the horizons:
$$
A^{tot}_n \sim 32\pi (n + {1\over 2})l_{Pl}^2 + 2A^{ext} + o(1).\eqno(3.27)
$$
As one can see, we have obtained a result which is closely related to the
proposal (1.3). However, it must be emphasized that Eq.(3.27) is not exactly
the same as Bekenstein's proposal (1.3): according to Bekenstein, the area of
the {\it outer} horizon of the hole has the spectrum (1.3), whereas we obtained
the same spectrum, with $\gamma=32\pi$, for the {\it total} area of the
horizons.
Nevertheless, it is most interesting that, according to our model, the
spectrum of the
total area of the horizons of the Reissner-Nordstr\"{o}m black hole is, up to
the term $2A^{ext}$, exactly the same as that of the Schwarzschild black hole.

           The WKB spectrum (3.20) has yet another property which is of some
interest. It follows from the mass formula of black holes that the ADM mass of
a non-rotating electrovacuum black hole is, from the point of view of a distant
observer at rest[33,34],
$$
M = {{\kappa}\over{4\pi}}A + \Phi Q.\eqno(3.28)
$$
In this equation, $\kappa$ is the surface gravity, $A$ is the area of the
(outer) horizon, and $\Phi$ is the
electric potential of the hole. For a Reissner-Nordstr\"{o}m black hole we have
$$\eqalignno{\kappa &= {{4\pi\sqrt{M^2-Q^2}}\over A},&(3.29.a)\cr
               \Phi &= {Q\over{r_{+}}}.&(3.29.b)\cr}
$$
Eq.(3.20) now implies that the WKB eigenvalues of the quantity
$$
{{\kappa}\over{4\pi}}A
$$
are of the form $\sqrt{2n}$. The physical interest of this result lies in its
charge independence. In other words, the
spectrum of the quantity $\kappa A\slash 4\pi$ is the same for the
Reissner-Nordstr\"{o}m and Schwarzschild black holes.

\bigskip

\bigskip

\centerline{\title 4. Quantum Theory with Charge as a}

\centerline{\title Dynamical Variable}

\bigskip

     In the previous Section we considered the electric charge of the
Reissner-Nordstr\"{o}m black hole as a mere external parameter, with no
dynamics whatsoever. In other words, we quantized only the gravitational degrees
of freedom of Reissner-Nordstr\"{o}m spacetimes. The object of this Section is
to extend our quantum theory of such spacetimes to include the electromagnetic
degrees of freedom as well. The guiding principle in our search for appropriate
canonical variables describing the dynamics of the electromagnetic field is
that since our distant obsever at rest observes the electromagnetic field
outside the event horizon as static, all the dynamics of the electromagnetic
field must be confined, classically, inside the horizon.

     To find  appropriate canonical variables, recall that when the classical
equations of motion are satisfied, the only non-vanishing component of the
electromagnetic potential $A_\mu$ is, in curvature coordinates, the component
$A_T=Q\slash R$. Now, this component is static with respect to the time $T$
everywhere outside the horizon. However, since $R$ becomes a timelike
coordinate when $r_-<R<r_+$, we find that $A_T$ necessarily has dynamics
between the inner and the outer horizons of the Reissner-Nordstr\"{o}m black
hole. In terms of $A_T$ we can write the Reissner-Nordstr\"{o}m metric (2.1)
as
$$
ds^2 = -\biggl(1 - {{2M}\over R} + A_T^2\biggr)\,dT^2 +
{{dR^2}\over{1-{{2M}\over R}+A_T^2}} + R^2\,d\Omega^2.\eqno(4.1)
$$

    In what follows, we shall "forget" the explicit dependence of $A_T$ on $R$
and $Q$, and instead treat $A_T$ as an independent dynamical variable of our
theory. However, in all our investigations we shall assume that $A_T$ is
independent of $T$ or, more precisely,
$$
{{\partial A_T}\over{\partial T}} \equiv 0.\eqno(4.2)
$$
>From this restriction it follows that we can treat $A_T$
as a function of an appropriate time coordinate only. Hence, our phase space,
which will be spanned by the throat variables $(a,p_a)$ and the electromagnetic
variables $(A_T,p_{A_T})$, where $p_{A_T}$ is the canonical momentum conjugate to
$A_T$, will be four-dimensional, which is in harmony with the results of
Ref.[5] reviewed in Section 2. To complete the classical theory we must just
find a canonical transformation from the "old" phase space variables
$(m,p_m,q,p_q)$ to the "new" phase space variables $(a,p_a,A_T,p_{A_T})$, and
write the classical Hamiltonian, from the point of view of our distant obsever
at rest, in terms of the variables $a,p_a,A_T$ and $p_{A_T}$.

      To find a clue to the expression of the classical Hamiltonian in terms of
the gravitational and electromagnetic variables, let us write the Hamiltonian
constraint on the timelike geodesic going through the bifurcation
two-sphere
of the Reissner-Nordtsr\"{o}m
spacetime, in the foliation used in Section 3. In that foliation the spacetime
metric can be written as
$$
ds^2 = -dt^2 + \biggl({{2M}\over a} - 1 - A_T^2\biggr)\,dT^2 +
a^2\,d\Omega^2.\eqno(4.3)
$$
As one can see, $T$ is now a spatial coordinate of spacetime. Hence, we can
identify the expression $({{2M}\over a} - 1 - A_T^2)^{1\slash 2}$ with the
variable $\Lambda(r,t)$ of Section 2. Moreover, we can identify $A_T$ with
$\Gamma$. The variable $\phi(r,t)$ is assumed to vanish. With these
identifications, and treating $M$ as a constant, we find that the Hamiltonian
contraint of Ref.[5] written in terms of $\Lambda$, $R$ and $\Gamma$ and their
time derivatives can be written in terms of $a$ and $A_T$ and their time
derivatives as
$$
{\cal H} = \biggl({{2M}\over a} - 1 - A_T^2\biggr)^{-{1\slash 2}}\biggl[{1\over
2}(1+A_T^2){\dot a}^2 + aA_T{\dot A}_T{\dot a} + {1\over 2}a^2{\dot A}_T^2 -
{M\over a} + {1\over 2}(1+A_T^2)\biggr] = 0.\eqno(4.4)
$$
>From this equation one can solve $M$:
$$
M= {1\over 2}a(1+A_T^2){\dot a}^2 + a^2A_T{\dot A}_T^2{\dot a} + {1\over
2}a(1+A_T^2).\eqno(4.5)
$$
It is easy to see that if one substitutes
$$
A_T = {Q\over a},\eqno(4.6)
$$
and keeps $Q$ as a constant, one gets
$$
M = {1\over 2}a{\dot a}^2 + {1\over 2}a + {{Q^2}\over{2a}}.\eqno(4.7)
$$
Hence, if one interprets the right hand side of Eq.(4.7) as the classical
Hamiltonian of the system, one gets, with the substitution (4.6), the same
Hamiltonian as in Eq.(3.7).

      At this point we define a new variable
$$
b := aA_T.\eqno(4.8)
$$
As a result, Eq.(4.5) becomes simplified to
$$
M = {1\over 2}a{\dot a}^2 + {1\over 2}a{\dot b}^2 + {1\over 2}a +
{{b^2}\over{2a}}.\eqno(4.9)
$$
Because of that, we are prompted to write the classical Hamiltonian of the
Reissner-Nordstr\"{o}m black hole, from the point of view of our distant
observer at rest, as
$$
H = {{p_a^2}\over{2a}} + {{p_b^2}\over{2a}} + {1\over 2}a +
{{b^2}\over{2a}},\eqno(4.10)
$$
where $p_a$ is, as in Section 3, the canonical momentum conjugate to the throat
radius $a$, and
$$
p_b := a{\dot b}\eqno(4.11)
$$
is the canonical momentum conjugate to the variable $b$.

    We obtained the Hamiltonian (4.10) by means of a guesswork based on the
study of the Hamiltonian constraint on the timelike geodesic going through the
bifurcation two-sphere, in a specific foliation of the Reissner-Nordstr\"{o}m
spacetime. The real problem is to find out, whether there exists a well defined
one-to-one canonical transformation from the phase space coordinates $m$, $p_m$,
$q$ and $p_q$, introduced in Section 2, to the phase space coordinates $a$,
$p_a$, $b$ and $p_b$ such that the Hamiltonian takes the form (4.10) if we
choose the lapse functions and the electric potentials at asymptotic infinities
as in Section 3.

    We shall perform such a transformation in two steps. We first define a
canonical momentum $p_w$ conjugate to a yet unknown variable $w$ as
$$
p_w := q.\eqno(4.12)
$$
With this choice the classical Hamiltonian takes the form
$$
H = {{p_a^2}\over{2a}} + {{p_w^2}\over{2a}} + {1\over 2}a.\eqno(4.13)
$$
The variables $m$ and $p_m$ are expressed in terms of $a$ and $p_a$ as in
Eqs.(3.4.a) and (3.6), but we have replaced $Q$ with $p_w$.

    The next task is to find $w$. One expects $w$ to be related in one way or
another to the momentum $p_q$ conjugate to $q$. Since $p_q$ defines the
electromagnetic gauge, we first write the Hamiltonian in a general gauge:
$$
H = {{p_a^2}\over{2a}} + {{p_w^2}\over{2a}} + {1\over 2}a + ({\tilde\phi}_+ -
{\tilde\phi}_-)p_w,\eqno(4.14)
$$
which follows from Eq.(2.35). Using Eq.(2.36.d) we find that the Hamiltonian
equation of motion for $w$ is
$$
{\dot w} = {{\partial H}\over{\partial p_w}} = {{p_w}\over a} - {\dot
p}_q.\eqno(4.15)
$$
An expression of $p_q$ in terms of $a$, $p_a$, $w$ and $p_w$ can be gained by
integrating the both sides of Eq.(4.15) along the classical trajectory in the
phase space:
$$
p_q = \int {{p_w}\over{a{\dot a}}}\,da - w,\eqno(4.16)
$$
where we have substituted
$$
{\dot a} = -sgn(p_w)\sqrt{{{2m}\over a} - 1 - {{p_w^2}\over a}}.\eqno(4.17)
$$
This substitution involves choosing ${\dot p}_q=0$. When the electric potential
is assumed to vanish at asymptotic infinities, this choice can be made. With an
appropriate choice of the integration constant we get
$$
p_q = sgn(p_m)p_w\biggl[\sin^{-1}\biggl({{p_a^2 + p_w^2
-a^2}\over{\sqrt{(p_a^2 + p_w^2 + a^2)^2 - 4a^2p_w^2}}}\biggr) +
{\pi\over 2}\biggr] - w,\eqno(4.18)
$$
where we have made the substitution
$$
m = {{p_a^2}\over{2a}} + {{p_w^2}\over{2a}} + {1\over 2}a.\eqno(4.19)
$$

     Eqs. (3.4), (4.12) and (4.18) now constitute a transformation from the
phase space coordinates $m$, $p_m$, $q$ and $p_q$ to the phase space
coordinates $a$, $p_a$, $w$ and $p_w$. It is easy to see that this
transformation is well defined and canonical. Moreover, the transformation is
one-to-one provided we impose a restriction
$$
\bigg\vert{{p_q + w}\over{p_w}}\bigg\vert \leq \pi.\eqno(4.20)
$$
This restriction is related to the fact that we are considering spacetime
between two successive $R=r_-$-hypersurfaces. Since ${\dot p}_q$ vanishes when
the electric potentials are assumed to vanish at asymptotic infinities, we find
that, classically, $w=-q\pi + p_q$ at the past $R=r_-$-hypersurface, $w=p_q$ at
the bifurcation point, and $w=q\pi + p_q$ at the future $R=r_-$ hypersurface.
In other words, the domain of $w$ is bounded by the fact that the
$t=constant$-hypersurfaces cannot be pushed beyond the $R=r_-$ hypersurfaces.

       It only remains to find a canonical transformation from the variables
$w$ and $p_w$ to the variables $b$ and $p_b$. We define
$$
\eqalignno{b &:= p_w\sin\biggl({w\over{p_w}}\biggr),&(4.21.a)\cr
          p_b&:= p_w\cos\biggl({w\over{p_w}}\biggr).&(4.21.b)\cr}
$$
This transformation is well defined and canonical as well as, with the
restriction (4.20), one-to-one. Because of that, we are justified to write the
classical Hamiltonian as in Eq.(4.10). Moreover, since it follows from
Eq.(4.21) that
$$
p_w^2 = p_b^2 + b^2,\eqno(4.22)
$$
we can identify the quantity $p_b^2 + b^2$ as the square of the electric charge
of the black hole.

     We now proceed to quantization. We choose our Hilbert space to be the
space $L^2(\Re^{+}\times\Re,a^s\,da\,db)$ with the inner product
$$
\langle\psi_1\vert\psi_2\rangle := \int_0^\infty da\,a^s\int_{-\infty}^\infty
db\,\psi^*_1(a,b)\psi_2(a,b).\eqno(4.23)
$$
Through the substitutions $p_a\rightarrow -i{\partial\over{\partial a}}$ and
$p_b\rightarrow -i{\partial\over{\partial b}}$ we replace the classical
Hamiltonian $H$ of Eq.(4.10) with the corresponding symmetric operator
$$
{\hat H} := -{1\over 2}a^{-s}{\partial\over{\partial
a}}\biggl(a^{s-1}{\partial\over{\partial a}}\biggr) -
{1\over{2a}}{{\partial^2}\over{\partial b^2}} + {1\over 2}a +
{{b^2}\over{2a}}.\eqno(4.24)
$$
As in Section 3, we can view the corresponding eigenvalue equation ${\hat
H}\psi=E\psi$ as an eigenvalue equation for the total ADM energy of the
Reissner-Nordstr\"{o}m black hole, from the point of view of a distant observer
at rest.

     The eigenvalue equation ${\hat H}\psi=E\psi$ can be separated if we write
$$
\psi(a,b) := \varphi(a)\beta(b),\eqno(4.25),
$$
and we get
$$
\eqalignno{\biggl[-{1\over 2}a^{-s}{d\over{da}}\biggl(a^{s-1}{d\over{da}}\biggr) + {1\over 2}a +
{{Q^2}\over{2a}}\biggr]\varphi(a) &= E\varphi(a),&(4.26.a)\cr
           \biggl(-{{d^2}\over{db^2}} + b^2\biggr)\beta(b) &=
Q^2\beta(b).&(4.26.b)\cr}
$$
Eq.(4.26.a) is now identical to Eq.(3.14), which is an eigenvalue equation for
the total energy of the hole when $Q$ is treated as an external parameter,
whereas Eq.(4.26.b) can be understood as an eigenvalue equation for the square
of the electric charge $Q$ of the hole. When the eigenfunctions $\beta(b)$ are
chosen to be the harmonic oscillator eigenfunctions, we find that the possible
eigenvalues of $Q^2$ are of the form
$$
Q_k^2 = 2k + 1,\eqno(4.27)
$$
or, in SI units:
$$
Q_k^2 = (2k+1){{e^2}\over{\alpha}},\eqno(4.28)
$$
where $k=0,1,2,3,...$. In this equation, $e$ is the elementary charge, and
$$
\alpha := {{e^2}\over{4\pi\epsilon_0\hbar c}}\approx {1\over{137}}\eqno(4.29)
$$
is the fine structure constant. In other words, our theory implies that the
electric charge of the Reissner-Nordstr\"{o}m black hole has a {\it discrete
spectrum}. It is interesting that the electric charge is quantized in terms of
the "Planck charge" $e\slash\sqrt{\alpha}$ in exactly the same way as the
quantity $\sqrt{M^2-Q^2}$ is quantized in terms of the Planck mass
$m_{Pl}:=(\hbar c\slash G)^{1\slash 2}$. Moreover, we find that the possible
eigenvalues of the area $A^{ext}:=4\pi Q^2$ of an extreme black hole are of the
form
$$
A^{ext}_k = 8\pi(k+{1\over 2})l_{Pl}^2.\eqno(4.30)
$$
In other words, we have recovered Bekenstein's proposal (1.3), with
$\gamma=8\pi$, for a black hole near extremality.
One can also see from Eq.(3.27) that even when the electric charge is made a
dynamical variable, the spectrum of the total area of the horizons
of the hole is the same as, according to Bekenstein, is the area spectrum of
the outer horizon of the hole.
However, it should be
noted that when $k$ is very big, the difference between two successive charge
eigenvalues is
$$
Q_{k+1} - Q_k \approx {{e^2}\over{\alpha Q_k}},\eqno(4.31)
$$
and we find that we have, in practice, a continuous charge spectrum. One may
therefore have very mixed feelings on the physical validity of the charge
spectrum (4.27): for all known particles --in appropriate units-- $Q$,
rather than $Q^2$, is an integer. However, for all these particles we have, in
natural units, $\vert Q\vert\gg M$, and nobody knows what happens when
$M\geq\vert Q\vert$.

\bigskip

\bigskip

\centerline{\title 5. Concluding Remarks}

\bigskip

            In this paper we have presented a quantum mechanical model of the
Reissner-Nordstr\"{o}m black hole, paying particular attention to the dynamical
properties of such black holes. More precisely, we used the throat radius of
the hole from the point of view of an observer in a radial free fall through
the bifurcation two-sphere of the hole, as the geometrodynamical variable of
our model. Motivated by the Principle of Equivalence, we chose the foliation of
spacetime in such a way that the proper time of a freely falling observer on
the throat agrees with the asymptotic Minkowski time. We then performed a
Hamiltonian quantization of our model. Our Hamiltonian quantum theory was based
on the analysis of the Hamiltonian dynamics of asymptotically flat spherically
symmetric electrovacuum spacetimes performed by Louko and Winters-Hilt.
According to that analysis, the true phase space of spacetimes mentioned above
is spanned by just four variables: the mass and the electric charge of the
Reissner-Nordstr\"{o}m hole, together with the corresponding canonical momenta. The
dynamical variables of our model --which in addition of gravitational, also
accounted for the electromagnetic degrees of freedom of the hole-- were
obtained by means of a canonical transformation from those used by Louko and
Winters-Hilt.

           Our model implied the results that the (ADM) mass and the electric
charge spectra of the Reissner-Nordstr\"{o}m black hole are discrete. Moreover,
we saw that the mass of the hole is bounded below and --with an appropriate
choice of the inner product and boundary conditions of the wave function of the
hole-- the mass
eigenvalues are always greater than the absolute values of the eigenvalues of
the electric charge of the hole. Hence, if we think of black hole radiation as
arising from a chain of transitions from higher to lower energy eigenstates of
the hole, our model implies that a non-extreme black hole with non-zero Hawking
temperature can never become, through black hole radiation, an extreme black
hole with zero Hawking temperature. This result is compatible with the third
law of thermodynamics as well as with the qualitative difference between extreme
and non-extreme black holes.

          At the high end of the mass spectrum, the WKB analysis yielded a
result that the possible eigenvalues of the {\it total} area of the outer and
the inner horizons of the hole are a constant plus an integer times $32\pi
l_{Pl}^2$, where $l_{Pl}$ is the Planck length. Hence, our result is closely
related, although not quite identical,
to Bekenstein's proposal (1.3). Moreover, we saw that no
matter whether the hole is charged or not, the spectrum of the total area of
its horizons is always essentially the same.

          Even though our model gives  a mathematically consistent quantum
theory of spherically symmetric, asymptotically flat electrovacuum spacetimes
and meets with
some success --in particular in its prediction that a non-extreme black hole
cannot become an extreme black hole by means of black hole radiation, and in its
close relationship with Bekenstein's proposal-- it is possible to express serious
objections against the physical relevance of our model.

          The first objection is related to the mass spectrum given by our
model. Recall that our model yields the result that the eigenvalues of the
quantity $\sqrt{M^2 -Q^2}$ are of the form $\sqrt{2n}$, where $n$ is an
integer. Now, this spectrum of the quantity $\sqrt{M^2-Q^2}$ implies that when
the hole performs a transition from a state with mass $M_{n+1}$ to the state
with mass $M_n := M$, the angular frequency of a quantum emitted in this
process is
$$
\omega \approx {1\over M}.\eqno(5.1)
$$
Regarding Hawking's results on black hole radiation, this kind of a spectrum
appears very natural when $Q=0$. In that case the expression (1.1) for Hawking
temperature, together with Wien's displacement law, implies that the maximum of
the black-body spectrum of the black hole radiation, as predicted by Hawking
and others, is proportional to ${1\slash M}$. In other words, the angular
frequency associated with the discrete spectrum of black hole radiation, as
predicted by our model, behaves, as a function of $M$, in the same way as does
the angular frequency corresponding to the maximum of the black-body spectrum
as predicted by Hawking and others. Unfortunately, this nice correspondence
between Hawking's results and our model breaks down when $Q\neq 0$. In that
case one finds from Eq.(1.1) that the maximum of the black-body spectrum
corresponds to the angular frequency
$$
\omega_{max} \propto {{\sqrt{M^2-Q^2}}\over{(M+\sqrt{M^2-Q^2})^2}}.\eqno(5.2)
$$
In other words, the angular frequency (5.1) predicted by our model corresponds,
when the hole is near extremality, to a temperature which is much {\it higher}
than the Hawking temperature.

         However, there is a possible way out from this problem. In all our
investigations we have emphasized the importance of the dynamics of the
intermediate region between the inner and the outer horizons of the
Reissner-Nordstr\"{o}m hole. The dynamics of the intermediate region is, in our
model, responsible for the discrete energy states of the hole. Now, if we take
this point of view to its extreme limits, we are prompted to speculate that it is the
intermediate region of the hole which emits black hole radiation when the hole
performs a transition from one energy eigenstate to another. Because of that,
it is possible that not only does the intermediate region emit radiation
outside the outer horizon but also {\it inside the inner horizon}. In other
words, both the inner and the outer horizons may radiate. The radiation emitted
by the inner horizon of the hole is not observed by an external obsever, and
presumably that radiation is either swallowed by the singularity or absorbed,
again, by the intermediate region. Nevertheless, an emission of this radiation by
the inner horizon is likely to reduce considerably the number of quanta, and
hence the temperature, of the radiation
coming out from the hole: the more the inner horizon radiates, the less
quanta are left for the outer
horizon. It remains to be seen whether this speculation of ours will be
validated by a further study.

           There is yet another, more fundamental, objection against our model.
In our model we have used a very small number of dynamical degrees of freedom,
together with the corresponding canonical momenta, to describe the quantum
mechanics of black holes. However, the fact that the Bekenstein-Hawking
entropy of a macroscopic black hole is very high suggests an enormous number of
dynamical degrees of freedom. Hence, one may question the relevance of models
with a small number degrees of freedom.

           However, there are some very powerful theorems on our side. They
are, of course, the black hole uniqueness theorems. According to these
theorems, a black hole in a stationary spacetime is uniquely characterized by
its mass, angular momentum and electric charge[35]. In other words, the number of
physical degrees of freedom of a black hole is, in the classical level, very
small.
Hence, we seem to have a slight disharmony between the Bekenstein-Hawking
entropy hypothesis and the black hole uniqueness theorems.

           A somewhat analogous situation can be met with already in ordinary
quantum mechanics. Consider a hydrogen atom. In elementary textbooks, the only
degrees of freedom under consideration are the three degrees of freedom
associated with the electron going around the proton. In more advanced
textbooks, however, a student is revealed that not only should one quantize the
degrees of freedom associated with the electron but also the degrees of freedom
associated with the electromagnetic field. As a result, one gets an enormous
number of degrees of freedom associated with the virtual photons and
electron-positron pairs appearing as an outcome of the quantization of the
electromagnetic field. In other words, although classically we have, in effect,
only the degrees of freedom associated with the electron, the full quantum
theory with quantized electromagnetic field reveals an enormous number of
particles and an enormous number of degrees of freedom. However, the whole
contribution of all these additional degrees of freedom to the energy levels of
the hydrogen atom is very small. Now, something similar may happen with black
holes: classically, the number of relevant degrees of freedom is very small, but
when the full quantum theory of gravity is employed, an enormous amount of
degrees of freedom are likely to appear. Hence, one may feel tempted to regard
the relationship between our model and the full quantum theory of black holes
as somewhat analogous to the relationship between the treatments of a hydrogen
atom in elementary and advanced textbooks: quantization of the two degrees of
freedom of the classical Reissner-Nordstr\"{o}m hole corresponds to the
quantization of the three degrees of freedom associated with the electron going
around the proton in a hydrogen atom. Whether the additional degrees of freedom
appearing as a likely outcome of the full quantum theory of black holes have
great or
small effects to the energy levels of the hole is an open question. However,
given the enormous pace of progress in the current research in black hole
physics, one may hope for a definite answer in a not so distant future.

\bigskip

\bigskip

\centerline{\title Acknowledgments}

\bigskip

     We thank Jorma Louko and Markku Lehto for their constructive criticism
during the preparation of this paper, and Matias Aunola for a numerical
investigation of Eq.(3.18).

\bigskip

\bigskip

\centerline{\title Appendix A: Solution of the Energy Eigenvalue}

\centerline{\title Equation for Large Energies and Charges}

\bigskip

In this Appendix we evaluate the large eigenvalues of the Hamiltonian
operator $\hat H$ which was written in Eq.(3.13). This leads us to the
eigenvalue equation (3.14) as we have already seen. We shall find the large
eigenvalue solutions of the eigenvalue equation by using the WKB approximation
method when both $\vert Q \vert$ and $E^2-Q^2$ are, in natural units, much
greater than unity and, in addition, we demand that $r_- \geq 1$ or, which is
the same thing, $(2E-1)/Q^2 \leq 1$.
The basic idea is to match the WKB-approximation with an expression of the
wave function in terms of modified Bessel functions close to the point where
$a=0$. The results used here on the matching of the WKB wave function and
Bessel function approximations close to the turning points are widely known --
see for example Ref.[34] -- therefore we shall use them without any special
review. Cases $r=1/2, \quad r \geq 3/2, \quad r=7/6, \quad 7/6<r<3/2$
and $1/2<r<7/6$ will be discussed separately, but we shall first look for the
general solution $\psi (a)$, when the argument $a$ is very small i.e.
$\vert Q \vert a \ll 1$. After that we shall search for the solution for
"slightly bigger" $a$ i.e $\vert Q\vert a \leq M$, where $M$ is an arbitrary
positive number.

To begin with we deform the eigenvalue equation (3.18) in an appropriate
manner. If we substitute in Eq.(3.18)
$$\eqalignno{a&:=x^{2/3},&(A1.a)\cr
             \chi &:= a^{-1/4}u(a),&(A1.b)\cr}$$
we get
$$ \biggl [{d^2\over da^2}-{({3\over 2}r-{1\over 4})({3\over 2}r-{5\over 4})
\over a^2}-a^2-Q^2+2Ea \biggl ] u(a) = 0,\eqno(A2)$$
where $r$ is defined in Eqs.(3.17). Eq.(A2) is now invariant under the
transformation $r\rightarrow1-r$; thus it is sufficient to consider solutions
of the eigenvalue equation (A2) for $r\geq1/2$. As a consequence the inner
product of Eq.(3.19) becomes
$$\langle u_1\vert u_2 \rangle = \int_0^\infty u_1^*(a) u_2(a)a\,da.\eqno(A3)$$

We shall now solve Eq.(A2) when $E^2-Q^2 \gg 1$, $\vert Q \vert \gg 1$ and
${(2E-1)/ Q^2 } \leq 1$. For very small $a$, the linearly independent
solutions to Eq.(A2) are, when $r >1/2$ and $r \neq 7/6$
$$\eqalignno{u_1(a) &= Aa^{(3/2)r}[a^{-1/4}+ {\cal O}(a^{7/4})],&(A4.a)\cr
            u_2(a) &= Ba^{-(3/2)r}[a^{5/4}+{\cal O}(a^{13/4})],&(A4.b)\cr}$$
where $A$ and $B$ are constants. The case $r=1/2$ will be considered later in
Appendix A, and if
$r = 7/6$ then the term proportional to $a^{-(3/2)r+13/4}$ in Eq.(A4.b) must be
multiplied by a term proportional to $\ln ({1\over 2}\vert Q \vert a)$. The
leading term, however, is the same as in Eq.(A4.b) when $r > 1/2$.

By writing
$$x=\vert Q\vert a,\eqno(A5)$$
we get from Eq.(A2) an equation
$$\biggl\lbrack{d^2\over dx^2}-{({3\over 2}r-{1\over 4})({3\over 2}r-{5\over 4})
\over x^2}-1-{x^2\over Q^4}+{2Ex\over \vert Q \vert ^3}\biggl\rbrack\,u(x) = 0.\eqno(A6)$$
Now the terms proportional to $x^2$ and $x$ are asymptotically small at large
$\vert Q \vert $, whenever $x \in (0,M]$, where $M$ is an arbitrary positive
constant. Omitting these last terms we get an equation
$$\biggl\lbrack{d^2\over da^2}-{({3\over 2}r-{1\over 4})({3\over 2}r-{5\over 4})
\over a^2}-Q^2\biggl\rbrack\,u(a) = 0,\eqno(A7)$$
when the substitution (A5) is inversed. The general linearly independent
solutions are, when ${3\over 2}(r-1/2)$ is not an integer, modified
Bessel functions of the first kind, up to an overall normalization constant
$$\eqalignno{u_1(a) &= a^{1/2}I_{{3\over 2}(r-1/2)}(\vert Q\vert a),&(A8.a)\cr
             u_2(a) &= a^{1/2}I_{-{3\over 2}(2-1/2)}(\vert Q \vert a).&(A8.b)\cr}$$
If ${3\over 2}(r-1/2)$ is an integer then the general solutions are, similarly
$$\eqalignno{u_1(a) &= a^{1/2}I_{{3\over 2}(r-1/2)}(\vert Q \vert a),&(A9.a)\cr
             u_2(a) &= a^{1/2}K_{{3\over 2}(r-1/2)}(\vert Q \vert a),&(A9.b)\cr}$$
where $K_p$ is the modified Bessel function of the second kind of order p.

\bigskip

\centerline{\it{1. Case $r\geq3/2$}}

\medskip

We first consider the case $r \geq 3/2$. Throughout the discussion we shall
assume that $E^2-Q^2>0$. The solution (A4) to Eq.(A2) is normalizable with
respect to the inner product (A3)
only if the constant B vanishes. Now, a comparison
with the asymptotic behaviour of the modified Bessel functions of Eq.(A8) for
small $a$ implies that the only normalizable solution for small $a$ is
$$u(a) = Ca^{1/2}I_{{3\over 2}(r-1/2)}(\vert Q \vert a)\eqno(A10)$$
when ${3\over 2}(r-1/2)$ is not an integer. If ${3\over 2}(r-1/2)$ is an
integer, a comparison
with Eq.(A9) gives similarly that the leading term is the same as in Eq.(A10).
To verify this, the Bessel functions must be expanded as their small $a$
series. If we fix $\delta_1$, $\delta_2 >0$ such that $\delta_1 \leq a \leq
\delta_2$, an asymptotic large $\vert Q \vert$ behaviour of $u(a)$ is, up to a
normalization constant:
$$u(a) \, \tilde{\propto}\, (2\pi \vert Q \vert )^{-1/2}\,
\exp ({\vert Q \vert a})\eqno(A11)$$
for all $r\geq 3/2$. From now on, the symbol $\tilde{\propto}$ is used for the
asymptotic form at large $\vert Q \vert$, up to a possibly $E,Q$--dependent
coefficient.

After very small, small, and slightly bigger argument $a$ we enter into a region
$a \in (0,\, a_- )$, where $a_-$ is the smaller turning point
satisfying for large $E$ and $\vert Q \vert$ $a_- \approx r_{-}$.
Our object is now to use the WKB approximation method to the wave function in
the region in question. The WKB approximation corresponding to such a wave
function $u(a)$ which decreases to the left of the
turning point $a_{-}$ is
$$u_{\hbox{WKB}}(a) = \lbrack p_1(a) \rbrack ^{-1/2}\exp \biggl[-\int_a^{a_-}p_1(a')da'
+ \eta_1\biggl],\eqno(A12)$$
where
$$p_1(a) = \sqrt{a^2+{(3/2r-1/4)(3/2r-5/4)\over a^2}-2Ea+Q^2}.\eqno(A13)$$
The major problem in the WKB approximation involves the evaluation of the
integral (A12). It turns out to be, however, that it is not necessary to evaluate the
WKB integral (A12) at all: we are interested in solutions for small $a$ i.e. $\vert
Q\vert a \ll 1$ and such solutions can be achieved easily by Taylor series.
In the evaluation of the series it should be clear that $E^2-Q^2$ and
$\vert Q\vert $ are assumed to be very large and ${(2E-1)/ Q^2} \leq 1$.
Furthermore, we assume that $a \geq \delta_1$ such that $\delta_1^2$ is
negligible. The integral in the exponent in Eq.(A12) is
$$S(a) := -\int_a^{a_-}\, da'\sqrt{a'^2+Q^2-2Ea'-{(3/2r-1/4)(3/2r-5/4)\over
a'^2}}.\eqno(A14)$$
With the help of Eq.(A14) the exponent for small $a$ in Eq.(A12) can be written
as
$$S'(\delta_1)(a-\delta_1)+{1\over 2}S''(\delta_1)(a-\delta_1)^2 + {\cal
O}(\delta_1^2),\eqno(A15)$$
where $S'$ denotes ${dS\over da}$.
Hence for small $a$, we have the WKB wave function given by Eqs.(A12) and
(A15)
$$u_{\hbox{WKB}}(a)\, \tilde{\propto}\, \vert Q \vert^{-1/2}
\exp (\vert Q \vert a),\eqno(A16)$$
where we have substituted the constant
$$\eta_1 = Q\delta_1.\eqno(A17)$$
This connects the WKB solution with the asymptotic solution in Eq.(A11). In
other words, the WKB solution decreases exponentially to the left of the turning
point $a_-$.

We next enter into an area with oscillations. We let the energy $E$ to be so large
that the eigenvalue equation has two turning points. We denote them as before
as $a_-$ and $a_+$. It should be clear that $a_- \approx r_-$ and $a_+ \approx
r_+$ when $E^2-Q^2$ and $\vert Q\vert$ are large enough. The region of oscillations is far right of $r_-$ and far
left of $r_+$. As the wave function decreases exponentially right of the
larger turning point and left of the smaller turning point, the WKB approximation
can be written far right  of $r_-$ as
$$u_{\hbox{WKB}}^{r_-}(a) = C_1\lbrack p_2(a)\rbrack^{-1/2}\cos \bigl\lbrack
\int_{a_-}^ada'p_2(a')-{\pi \over 4}\bigl\rbrack,\eqno(A18.a)$$
and far left of $r_+$ as
$$u_{\hbox{WKB}}^{r_+}(a) = C_1\lbrack p_2(a)\rbrack ^{-1/2} \cos \bigl\lbrack
\int_{a} ^{a_+} da'p_2(a')-{\pi \over 4}\bigl\rbrack,\eqno(A18.b)$$
where
$$p_2(a) :=
\sqrt{-a^2+2Ea-Q^2}\sqrt{1-{(3/2r-1/4)(3/2r-5/4)\over a'^2(-a'^2-Q^2+2Ea')}}.
\eqno(A19)$$
The wave functions above are equal if
$$S:=\int_{a_-}^{a_+}da\,p_2(a) = (n+{1\over 2})\pi,\eqno(A20)$$
where $n \geq 0$ is an integer. This integral fixes the levels of the spectrum
of the Reissner--Nordstr\"om black hole.

In the evaluation of the WKB integral it should, again, be clear that
$E^2-Q^2$ and $\vert Q\vert$ are assumed very large.
We can expand the second square root in its Taylor series as the first square
root is of order  ${\cal O}(E^2-Q^2)$ and the second term in the second square
root is of order  ${\cal O}(1/({a_-}^2\sqrt{E^2-Q^2}))$.
Thus we can write the integral as
$$S = \int_{a_-} ^{a_+} \,
da\Biggl [ \sqrt{-a^2-Q^2+2Ea}-
{(3/2r-1/4)(3/2r-5/4)\over 2a^2\sqrt{-a^2-Q^2+2Ea}} - \ldots \Biggl].\eqno(A21)
$$
Now that $a_- = r_- +{\cal O} (1/{r_-}^3)$ and $a_+ = r_+ - {\cal O}(1/{r_+}^3)
$, the evaluation of the integral $S$ can be done in parts, and by replacing
the limits $a_-$ and $a_+$ by $r_-$ and $r_+$
the second integral gives us a term order of ${\cal O}(E/Q^3)$, which, on the grounds of
the requirement $(2E-1)/Q^2 \leq 1$, is small compared to
the first term. Thus the second term can be omitted from $S$.
In the evaluation of the third term from $r_-$ to $r_+$ the integral does not
converge. We therefore have to alter the integration region near the turning
points by fixing  a couple of constants, namely,
$\delta_3,\, \delta_4 >0$ such that we are able
to restrict the argument $a$ in the region $a_- < r_- +\delta_3/(E^2-Q^2)^{1/8}
\leq a \leq r_+ -\delta_4/(E^2-Q^2)^{1/8} <a_+$ for large enough $E^2-Q^2$.
Then the limits can be replaced by $r_- +\delta_3/(E^2-Q^2)^{1/8}$ and $r_+ -
\delta_4/(E^2-Q^2)^{1/8}$. Now the third term
in the integral $S$ gives us a term, which is at most of order ${\cal O}({r_-}^{-4}(E^2-Q^2)^{-1/4})$,
which is small compared to the first term and can thus be omitted. The
remaining integral is elementary and we obtain
$$S \approx \int_{r_-}^{r_+}da\,\sqrt{-a^2+2Ea-Q^2} = {\pi \over 2}(E^2-Q^2) =
(n+{1\over 2})\pi.\eqno(A22)$$
This yields for large energies and charges, when $r \geq 3/2$ and $r_- \geq 1$,
the WKB estimate
$$E^2-Q^2 \sim 2n +1 + o(1).\eqno(A23)$$

\bigskip

\centerline{\it{2. Cases $7/6 <r<3/2$ and $1/2<r<7/6$}}

\medskip

Now we are in a situation where we cannot just exclude either of the integration
constants $A$ or $B$ in Eq.(A4) on the grounds of the normalizability of the wave function.
However, the self--adjointness of the Hamiltonian operator implies the
following boundary condition for the solutions $u_{1,2}(a)$:
$$\lim _{a\to 0}\biggl\lbrack u_1^*(a){du_2(a) \over da} - {du_1^*(a) \over da}
u_2(a)\biggl\rbrack = 0.\eqno(A24)$$
Here $u_1$ and $u_2$ are two linearly independent, non--degenerate eigenfunctions.
As shown in Eq.(A4) the differential equation $\hat H u(a) = E u(a)$ has two
small $a$ solutions which satisfy all those conditions stated above -- at least when $E^2
-Q^2>0$. Now it is easy to show that, for very small a, the eigenfunctions of a self--adjoint Hamiltonian
operator behave, up to normalization, as
$$u(a) \approx \cos (\theta)a^{(3/2)r-1/4}+\sin (\theta)a^{-(3/2)r+5/4},\eqno(A25)$$
where $\theta \in \lbrack 0,\pi)$ is a parameter to be fixed later.
Comparing the small $a$ expansions of Eqs.(A25) and (A8) we can adjust the
constants $A$ and $B$ such that $u(a)$ behaves asymptotically
$$\eqalignno{u(a) \, \tilde{\propto} \, &a^{1/2}\biggl\lbrack
2^{{3\over 2}(r-1/2)}\Gamma (3/2r+3/4)\vert Q \vert ^{-{3\over2}(r-1/2)}
\cos (\theta)I_{{3 \over 2}(r-1/2)}(\vert Q \vert a)&\cr
&+ 2^{-{3\over 2}(r-1/2)}\Gamma(-3/2r+5/4)\vert Q \vert ^{{3\over 2}(r-1/2)}
\sin (\theta)I_{-{3\over 2}(r-1/2)}(\vert Q \vert a)
\biggl\rbrack&.(A26)\cr}$$
When $\theta =0$, the second term in Eq.(A26) vanishes and we can proceed just as
in the case $r \geq 3/2$ from which it follows that the WKB estimate is
given by Eq.(A23). When $\theta \neq 0$
the second term in Eq.(A26) dominates at large $\vert Q \vert$ and the asymptotic
behaviour is as Eq.(A11). Therefore the WKB estimate is again
given by Eq.(A23).

\bigskip

\centerline{\it{3. Case $r = 7/6$}}

\medskip

When $r=7/6$, the number ${3\over 2}(r-1/2)$ becomes an integer and the general solution of
Eq.(A7) includes modified Bessel functions of the second kind as shown before.
Furthermore, we cannot rule out either of the adjustable constants $A$ or $B$ and
therefore we have to keep both the solutions in Eq.(A9). As before
we get from the boundary condition (A24) that at least when $E^2-Q^2>0$ the
eigenfunction of a self--adjoint Hamiltonian operator is, for small $a$,
$$u(a) \approx \sin (\theta)a^{-1/2}+\cos (\theta)a^{3/2},\eqno(A27)$$
where again $\theta \in \lbrack 0,\pi )$ is a parameter.

After expanding the general solution of Eq.(A7) when $a$ is small we have that
$u(a)$ is asymptotically
$$u(a) \,\tilde{\propto} \, a^{1/2}\bigl\lbrack
(2\cos (\theta)\vert Q \vert ^{-1}-\sin (\theta)\vert Q \vert \gamma)I_1
(\vert Q\vert a)+\sin (\theta)\vert Q \vert K_1(\vert Q \vert a)
\bigl\rbrack,\eqno(A28)$$
where $\gamma$ is Euler's constant.
When $\theta = 0$ the term proportional $a^{1/2}K_1(\vert Q \vert a)$ vanishes and we get
the same WKB estimate as before in Eq.(A23). On the other hand, when
$\theta \neq 0$, the term proportional $a^{1/2}I_1(\vert Q \vert a)$ dominates for
large $\vert Q \vert$ and the situation is quite the same as before. The
resulting WKB estimate is therefore given by Eq.(A23).

\bigskip

\centerline{\it{4. Case $r=1/2$}}

\medskip

When $r=1/2$ we can no more write the solutions of Eq.(A7) as powers of small
$a$,
because the loss of the linear independency of the solutions (A4.a) and (A4.b). By using the boundary
condition (A24), however, and expanding the general solution $u(a) =
a^{1/2}\lbrack
C\,I_0(\vert Q\vert a)+D\,K_0(\vert Q\vert a)\rbrack$ of Eq.(A7) for small $a$
we notice that Eq.(A26) is replaced by
$$u(a) \, \tilde{\propto} \, \vert Q\vert^{1/4}a^{1/2}\lbrack
(\cos (\theta)-\sin (\theta)(\gamma +\ln ({1\over 2}\vert Q\vert)))I_0
(\vert Q \vert a)
- \sin (\theta)K_0(\vert Q \vert a) \rbrack.\eqno(A29)$$
For any $\theta$ the term proportional to $a^{1/2}I_0(\vert Q \vert a)$ dominates the term
proportional to $a^{1/2}K_0(\vert Q\vert a)$ for large charges, and the asymptotic behaviour is
given by Eq.(A11). Thus the WKB result is, again, given by Eq.(A23).

It should be noted that we have not investigated what happens when the
condition $(2E-1)/Q^2 \leq 1$ following from the requirement $r_- \geq 1$ does
not hold; i.e. when $Q$ is arbitrarily small
when compared to $E$. In that case, however, one expects that the WKB
eigenenergies given by Eq.(A23) should be replaced by those given in Ref.[8] for the
Schwarzschild black hole. On the grounds of the results of Ref.[8] it is
likely that the eigenvalues of
the quantity $\sqrt{E^2-Q^2}$ are of the form $\sqrt{2n}$ even when $\vert Q
\vert$ is arbitrarily small compared to the black hole energy $E$.

\bigskip

\bigskip

\centerline{\title Appendix B: Positiveness of the Spectrum of}

\centerline{\title the Quantity $E^2-Q^2$}

\bigskip

In this appendix we shall investigate the possible positiveness of the spectrum
of the quantity $E^2-Q^2$. Cases $r=1/2$ and $r \geq 3/2$ and $1/2 <r< 3/2$
will be considered separately.

\bigskip

\centerline{\it{1. Case $r\geq 3/2$}}

\medskip

Let $u(a)$ be an eigenfunction of Eq.(A2) with any eigenvalue
$E^2-Q^2$. Now, when $r\geq 3/2$, we have, up to a well chosen normalization
constant, a small $a$ expansion to the eigenfunction $u(a)$ given by Eq.(A4.a)
as $u(a) = a^{(3/2)r}[a^{-{1\over 4}}+{\cal O}(a^{7/4})].$
It is clear that $u(a)$  and $u'(a)$ are both real valued and positive. Therefore
the eigenfunction is positive and real valued for sufficiently small $a$.
It is easy to see that the eigenvalue equation (A2) can be written as
$$u''(a) = \Biggl[{ ({3\over 2}r-{1\over 4})({3\over 2}r-{5\over 4})\over a^2} +
(a-E)^2 -(E^2-Q^2)\Biggl]u(a).\eqno(B1)$$
Let us now assume $E^2-Q^2$ is not strictly positive i.e. $E^2-Q^2 \leq 0$. In
that case Eq.(B1) implies that
$u''(a)>0,$
for all $a$ such that $u(a) >0$. Since both $u(a)$ and $u'(a)$ are positive for
sufficiently small $a$, the positivity of $u''(a)$ whenever $u(a)$ is positive
implies that $u'(a)$, and hence $u(a)$, are increasing functions of $a$.
Because of that, we have $\lim_{a\rightarrow \infty} u(a) > 0$ and $u(a)$ is not normalizable. Hence
we must have $E^2-Q^2 >0$.

\bigskip

\centerline{\it{2. Case $1/2<r<3/2$}}

\medskip

Here we shall show that
it is possible to find such self--adjoint extensions of the
Hamiltonian operator such that the spectrum is strictly positive. This
corresponds to an appropriate
choice of the parameter $\theta$ introduced in Appendix A in Eq.(A25). We have
already shown that  when $7/6<r<3/2$ and $1/2<r<7/6$ the extensions take, up
to an overall normalization constant, for small $a$ the form given in Eq.(A25), where
the parameter $\theta$ specifies the self--adjoint extensions.

We consider first an extension with $\theta \in [0,\pi /2]$.
By taking $E^2-Q^2$ in Eq.(B1) to be negative or zero, we get, when $u(a)$ is
positive,
$$u''(a) \geq {({3\over 2}r-{1\over 4})({3\over 2}r-{5\over 4})\over
a^2}u(a) \ \ \forall a>0.\eqno(B2)$$
To find a possible lower bound for $u(a)$ we therefore consider an equation
$$f''(a) = {({3\over 2}r-{1\over 4})({3\over 2}r-{5\over 4})\over
a^2}f(a).\eqno(B3)$$
Now when $r>1/2$ the general solution of Eq.(B3) is
$$f(a) = Aa^{(3/2)r-1/4}+Ba^{-(3/2)r+5/4},\eqno(B4)$$
and we see that if we choose $A=\cos \theta$ and $B=\sin \theta$ then $f(a)$
coincides with the small $a$ solution (A25).
Since $u(a)$ is positive for sufficiently small $a$ when both $\cos \theta$ and
$\sin \theta$ are positive, we find that the
solution $u(a)$ is equal or greater than the solution $f(a)$ for all $a\geq 0$
when $0 \leq \theta \leq \pi/2$. Hence the solution $u(a)$ does not vanish
exponentially as $a$ goes to infinity. Yet, the potential increases without bound
as $a$ increases towards infinity; therefore any eigenfunction must vanish exponentially
at large $a$. Hence the spectrum must be strictly positive.

As to the remaining range $\theta \in (\pi /2, \pi )$  we refer to the results
of Ref.[8], which state that the energy spectrum is bounded below.

\bigskip

\centerline{\it{3. Case $r=1/2$}}

\medskip

For  $r=1/2$ Eqs.(B2) and (B3) still hold and Eq.(B4) must be replaced by a solution
$$f(a) = (\cos \theta -\sin \theta )a^{1/2} -\sin \theta a^{1/2}\ln a.\eqno(B5)$$
When $\theta = 0$, $f(a)$ is positive for all $a\geq 0$ and one can argue as before.

In the remaining range $0<\theta <\pi $ the energy spectrum is bounded below on the
grounds of the results presented in Ref.[8].

\vfill\eject

\centerline{\title References}

\bigskip

[1] J. B. Hartle and S. W. Hawking, Phys. Rev. {\bf D13}, 2188 (1976)

\medskip

[2] S. W. Hawking, G. T. Horowitz and S. F. Ross, Phys. Rev. {\bf D51}, 4302

(1995)

\medskip

[3] C. Teitelboim, Phys. Rev. {\bf D51}, 4315 (1995)

\medskip

[4] T. Brotz and C. Kiefer, Phys. Rev. {\bf D55}, 2186 (1997)

\medskip

[5] J. Louko and S. N. Winters-Hilt, Phys. Rev. {\bf D54}, 2647 (1996)

\medskip

[6] S. W. Hawking and G. F. R. Ellis, {\it The Large Scale Structure of
Space-time}

(Cambridge University Press, Cambridge, England, 1973)

\medskip

[7] C. W. Misner, K. S. Thorne and J. A. Wheeler, {\it Gravitation} (Freeman,
San

Francisco, 1973)

\medskip

[8] J. Louko and J. M\"{a}kel\"{a}, Phys. Rev. {\bf D54}, 4982 (1996)

\medskip

[9] J. D. Bekenstein, Lett. Nuovo Cimento {\bf 11}, 467 (1974)

\medskip

[10] K. Kucha\v{r}, Phys. Rev. {\bf D50}, 3961 (1994)

\medskip

[11] C. Rovelli, Int. J. of Theor. Phys. {\bf 35}, 1637 (1996)

\medskip

[12] V. F. Mukhanov, Pis'ma Zh.Eksp.Teor.Fiz. {\bf 44}, 50 (1986) (JETP Lett.

{\bf 44}, 63 (1986))

\medskip

[13] I. Kogan, Pis'ma Zh. Eksp. Teor. Fiz. {\bf 44}, 209 (1986)

\medskip

[14] P. O. Mazur, Phys. Rev. Lett. {\bf 57}, 929 and {\bf 59}, 2380 (1987)

\medskip

[15] P. O. Mazur, Gen. Rel. Grav. {\bf 19}, 1173 (1987)

\medskip

[16] V. F. Mukhanov, in {\it Complexity, Entropy and the Physics of
Information},

ed. by W. H. Zurek, Vol. VIII (Addison-Wesley Publ. Comp. Redwood

City, California 1990)

\medskip

[17] J. Garcia-Bellido, Report SU-ITP-93-4, hep-th/9302127

\medskip

[18] U. H. Danielsen and M. Schiffer, Phys. Rev. {\bf D48}, 4779 (1993)

\medskip

[19] Y. Peleg, Report BRX-TH-350, hep-th/9412232

\medskip

[20] M. Maggiore, Nucl. Phys. {\bf B429}, 205 (1994)

\medskip

[21] I. Kogan, Report OUTP-94-39P, hep-th/9412232

\medskip

[22] C. O. Lousto, Phys. Rev. {\bf D51}, 1733 (1995)

\medskip

[23] Y. Peleg, Phys. Lett. {\bf B356}, 462 (1995)

\medskip

[24] J. D. Bekenstein and V. F. Mukhanov, Phys. Lett. {\bf B360}, 7 (1995)

\medskip

[25] V. Berezin, Phys. Rev. {\bf D55}, 2139 (1997)

\medskip

[26] P. O. Mazur, Acta Phys. Polon. {\bf 27}, 1849 (1996)

\medskip

[27] H. A. Kastrup, Phys. Lett. {\bf B385}, 75 (1996)

\medskip

[28] A. Barvinskii and G. Kunstatter, Phys. Lett. {\bf B329}, 231 (1996)

\medskip

[29] J. M\"{a}kel\"{a}, Phys. Lett. {\bf B390}, 115 (1997)

\medskip

[30] V. Berezin, gr-qc/9701017

\medskip

[31] A. Z. Gorski and P. W. Mazur, hep-th/9704179

\medskip

[32] H. A. Kastrup, gr-qc/9707009

\medskip

[33] See, for example, I. D. Novikov and V. P. Frolov, {\it Physics of Balck
Holes}

(Kluwer Academic Publishers, Dortrecht, Holland, 1989), and

\medskip

[34] B. Carter in {\it General Relativity, An Einstein Centenary Survey},

edited  by S. W. Hawking and W. Israel (Cambridge University Press, New

York, 1979)

\medskip

[35] See, for example, M. Heusler, {\it Black Hole Uniqueness Theorems}

(Cambridge University Press, Cambridge, England, 1996)

\medskip

[36] A. Messiah, {\it Quantum Mechanics, Vol. 1} (Noth-Holland Publ. Comp.,

Amsterdam, 1991)

\bye